\documentclass{article}
\usepackage{graphicx}  
\usepackage{amsmath}   
\usepackage[compress]{cite}
\usepackage{amssymb}   
\usepackage{bm} 
\usepackage{dcolumn}
\usepackage{color}
\usepackage{mathrsfs}
\usepackage{amsfonts}
\usepackage[utf8]{inputenc}
\usepackage{varioref}
\usepackage{csquotes}
\usepackage[caption=false]{subfig}
\RequirePackage[colorlinks,citecolor=blue,urlcolor=magenta,linkcolor=blue]{hyperref}
\def\KNN{Kerr-Newman-NUT}
\def\CPP{collisional Penrose process}
\def\ER{ergoregion}
\def\N{NUT}

\def\SP{superradiance}
\labelformat{section}{Section #1} 
\labelformat{subsection}{Section #1} 
\labelformat{subsubsection}{Section #1}
\labelformat{subsubsubsection}{Section #1}
\labelformat{equation}{Eq.~(#1)} 
\labelformat{figure}{Fig.~#1} 
\labelformat{subfigure}{Fig.~\thefigure#1} 
\labelformat{table}{Table~#1} 
\labelformat{appendix}{Appendix #1}
\title{On some novel features of the Kerr-Newman-NUT Spacetime}
\author{Sajal Mukherjee\footnote{sm13ip029@iiserkol.ac.in}$~^{1}$, 
Sumanta Chakraborty\footnote{sumantac.physics@gmail.com}$~^{2}$ 
and 
Naresh Dadhich\footnote{nkd@iucaa.in}$~^{3}$\\
{$^{1}$\small{Department of Physical Sciences, Indian Institute of Science Education and Research, Kolkata-741246, India}}\\
{$^{2}$\small{School of Physical Sciences, Indian Association for the Cultivation of Science, Kolkata-700032, India}}\\
{$^{3}$\small{Inter-University Centre for Astronomy and Astrophysics, Post Bag 4, Pune-411007, India}}}
\begin{document}
  
\maketitle
\begin{abstract}
In this work we have presented a special class of Kerr-Newman-NUT black hole, having its horizon located precisely at $r=2M$, for $Q^{2}=l^{2}-a^{2}$, where $M$, $l$, $a$ and $Q$ are respectively  mass, NUT, rotation and electric charge parameters of the black hole. Clearly this choice radically alters the causal structure as there exists no Cauchy horizon indicating spacelike nature of the singularity when it exists. On the other hand, there is no curvature singularity for $l^2 > a^2$, however it may have conical singularities. Furthermore there is no upper bound on specific rotation parameter $a/M$, which could exceed unity without risking destruction of the horizon. To bring out various discerning features of this special member of the Kerr-Newman-NUT family, we study timelike and null geodesics in the equatorial as well as off the equatorial plane, energy extraction through super-radiance and Penrose process, thermodynamical properties and also the quasi-periodic oscillations. It turns out that the black hole under study radiates less energy through the super-radiant modes and Penrose process than the other black holes in this family.
\end{abstract}
\section{Introduction}

Duality of Maxwell's equations in the presence of magnetic monopole has far reaching consequences. Therefore, it seems legitimate to explore whether there can exist any such duality for gravitational dynamics as well. Surprisingly, it turns out that there is indeed such a duality in the realm of gravitational field of a Kerr-Newman black hole. For asymptotically flat spacetimes, the unique solutions of the Einstein-Maxwell field equations are the black holes in Kerr-Newman family \cite{Newman:1965my}. However if the condition of asymptotic flatness is dropped, then one can have an additional hair on the black hole, known as the NUT charge and the black holes are referred to as the Kerr-NUT solutions \cite{Newman:1963yy,demianski1966combined}. We also refer our readers to Refs. \cite{Chen:2006xh,Kubiznak:2009ad,Awad:2005ff,Dehghani:2005zm,Mukhopadhyay:2003iz} for a descriptive overview on NUT solutions and their various implications in Einstein as well as alternative theories of gravity. Besides admitting separable Hamilton-Jacobi and Klein-Gordon equations \cite{Dadhich:2001sz}, the above family also shares a very intriguing duality property: the spacetime structure is invariant under the transformation $\textrm{mass}\leftrightarrow \textrm{NUT charge}$ and $\textrm{radius}\leftrightarrow \textrm{angular coordinate}$ \cite{Turakulov:2001jc}. Given this duality transformation one can associate a physical significance to the NUT parameter, namely a measure of gravitational magnetic charge. Even though it is possible to arrive at a NUT solution without having rotation, the above duality only works if the rotation parameter is non-zero \cite{Argurio:2009xr}. Thus in order to have a concrete theoretical understanding of the present scenario rotation is necessary. 

On the other hand, even though there is no observational evidence whatsoever for the existence of gravitomagnetic mass \cite{LyndenBell:1996xj}, investigation of the geodesics in Kerr-Newman-NUT spacetime has significance from both theoretical as well as conceptual points of view. The observational avenues to search for the gravitomagnetic monopole includes, understanding the spectra of supernovae, quasars and active galactic nuclei \cite{LyndenBell:1996xj}. All of these scenarios require presence of thin accretion disk \cite{GarciaReyes:2004qn} and can be modelled if circular geodesics in the spacetime are known. Thus a proper understanding of the geodesic motion in the presence of NUT parameter is essential. Following such implications in mind, there have been attempts to study circular timelike geodesics in presence of NUT parameter \cite{Chakraborty:2013kza,Pradhan:2014zia,Jefremov:2016dpi}\footnote{Note that the results presented in \cite{Chakraborty:2013kza,Pradhan:2014zia} are based on the erroneous assumption that circular geodesics lie on the equatorial plane, see e.g., \cite{Jefremov:2016dpi}.} as well as motion of charged particles in this spacetime \cite{Cebeci:2015fie}. Various weak field tests, e.g., perihelion precession, Lense-Thirring effect has also been discussed \cite{Chakraborty:2012wv,Chakraborty:2013naa,Chakraborty:2014jaa} (for a taste of these weak field tests in theories beyond general relativity, see \cite{Chakraborty:2011uj,Chakraborty:2012sd,Chakraborty:2015vla,Bhattacharya:2016naa,Mukherjee:2017fqz}). We would like to emphasize that most of these studies on the geodesic motion crucially hinges on the equatorial plane, however there can also be interesting phenomenon when off-the-equatorial plane motion is considered. 

Besides understanding the geodesic structure, it is of utmost importance to explore the possibilities of energy extraction from black holes as well. In particular, the phenomenon of Penrose process \cite{penrose1971extraction}, superradiance \cite{zel1971generation,Starobinsky:1973aij} and the Ba\~{n}ados-Silk-West effect are well studied in the context of Kerr spacetime \cite{Banados:2009pr}. Implications and modifications to these energy extraction processes in  presence of NUT charge is another important issue to address. It will be interesting to see how the efficiency of energy extraction in the Penrose process depends on the gravitomagnetic charge inherited by the spacetime. Further, being asymptotically non-flat, whether some non-trivial corrections to the energy extraction process appear is something to wonder about. Moreover, it is expected that the phenomenon of superradiance and the counter-intuitive Banados-Silk-West effect will inherit modifications over and above the Kerr spacetime due to presence of the NUT charge. In particular, for what values of angular momentum the center-of-mass energy of a system of particles diverges is an interesting question in itself.

The uniqueness theorems dictates that a black hole can have only three hairs, mass, angular momentum and electric charge in an asymptotically flat spacetime. If the requirement of asymptotic flatness is relaxed, it can have a new hair, namely the NUT parameter. Each of these black hole hairs must be tested on the anvil of astrophysical observations. So the question is, could we work out an observationally testable effect that could put bounds on NUT paramater. For that we have studied quasi periodic oscillations for the black hole in question where fundamental frequency of oscillations depends upon it. This could be one of the possible observational tests to unveil the existence of NUT parameter.  

Finally we should say a word about our choice of the metric for this investigation. In the Kerr-Newman-NUT metric, rotation, NUT charge and the electric charge parameters, $a^2$, $l^2$ and $Q^2$, appear linearly in $\Delta = r^2-2Mr-l^2+a^2+Q^2$. Thus if we make the following unusual choice: $Q^2+a^{2}=l^2$ \cite{Dadhich:2002bb}\footnote{Since it is well known that presence of rotation also produces gravomagnetic effects, it is not out of place to consider a relation between them.}, $\Delta$ becomes simply $r^2-2Mr$ and thereby black hole horizon is entirely determined by mass alone and coincides with that of the Schwarzschild's. Despite this the black hole is having both electric and NUT charge and also rotating. This happens because the electric charge appears only in $\Delta$ and nowhere else in the metric, while NUT and rotation parameters also define geometrical symmetry of the spacetime. This is why it could be simply added or subtracted, i.e., in $\Delta$ of Kerr-NUT metric, simply add $Q^2$ to obtain the Kerr-Newman-NUT solution of the Einstein-Maxwell equations. It is noteworthy that despite presence of rotation and electric charge, the singularity is not timelike but rather spacelike, as of the Schwarzschild black hole. That is, the choice of $Q^2+a^2=l^2$ indicates that repulsive effect due to charge and rotation is fully balanced by attractive effect due to the NUT parameter. This is why the causal structure of spacetime has been radically altered \cite{hawking1973large}. Unlike any other black holes in this family, it would have spacelike singularity when it exists. For existence of singularity, one must have  $r^2 + (l + a \cos\theta)^2 = 0$, which will never be so for $l>a$. Thus spacelike singularity will only exist for $l \leq a$, else for $l>a$ it would be free of the ring singularity at $r=0$. Another remarkable feature is that the specific rotation parameter $a/M$ could have any values even exceeding unity without risking the singularity turning naked. All these are very novel and interesting features, and their exposition is the main aim of this paper. With the choice $l>a$, there occurs no ring singularity as reflected in the fact that $r^2 + (l + a \cos\theta)^2 \neq 0$ for any choices of $r$ and $\theta$. It is therefore a very interesting special case of the Kerr-Newman-NUT family of spacetimes, whose structure we wish to understand in this paper for studying its various interesting properties.

The paper is organized as follows: In \ref{KNN_Spacetime} we have elaborated the spacetime structure we will be considering in this work. Subsequently in \ref{KNN_Trajectory} the trajectory of a particle in both equatorial and non-equatorial plane has been presented. Various energy extraction processes in this spacetime have been illustrated in \ref{Energy_KNN} and finally thermodynamics of black holes in presence of NUT charge has been jotted down in \ref{KNN_Thermo}. We finish with a discussion on the results obtained in \ref{KNN_Conc}. 

\textit{Notations and Conventions:} Throughout this paper we have set the fundamental constants $c=1=G$. All the Greek indices run over four dimensional spacetime coordinates, while the roman indices run over spatial three dimensional coordinates. 
\section{The spacetime structure}
\label{KNN_Spacetime}

The Kerr-Newman-NUT solution in general involves  mass of the black hole $M$, rotation parameter $a$, electric charge $Q$ and NUT charge $l$ and the spacetime metric is given by  
\begin{align}\label{metric_KN_old}
ds^{2}=-\frac{\Delta}{\rho ^{2}}\left(dt-Pd\phi\right)^{2}
+\frac{\sin ^{2}\theta}{\rho ^{2}}\left\{\left(r^{2}+a^{2}+l^{2}\right)d\phi -adt\right\}^{2}
+\frac{\rho ^{2}}{\Delta}dr^{2}+\rho ^{2}d\theta ^{2} .
\end{align}
where the quantities $\Delta$, $P$ and $\rho ^{2}$ have the following explicit expressions,
\begin{align}\label{def_eq}
\Delta = r^{2}-2Mr+a^{2}+Q^{2}-l^{2};\qquad P=a\sin ^{2}\theta -2l\cos \theta;\qquad \rho ^{2}=r^{2}+(l+a\cos \theta)^{2} .
\end{align}
As evident for $l=0$, we get back the Kerr-Newman solution, while for $Q=0$ it is the Kerr-NUT solution (for a more detailed discussion on the \KNN\ solution, see \cite{Griffiths:2009dfa,Grenzebach:2014fha,Johnson:2014pwa}). The location of the horizons can be obtained by solving for $\Delta=0$, which in general will have two distinct roots and this is in contrast to the one horizon of Schwarzschild black hole. Furthermore, the curvature singularity in this context is timelike and located at $\rho^{2}=0$. This corresponds to a ring having $r=0$ and $\cos \theta=-\ell/a$. That means curvature singularity will only occur for $l \leq a$ else for $l>a$, spacetime will be free of curvature singularity. However, this does not warrant that the spacetime will be regular everywhere. In particlar, presence of the NUT charge generates a conical singularity on its axis of symmetry having poles at $\theta=0$ and $\theta=\pi$. It is possible to get rid off conical singularities by imposing a periodicity condition over the time coordinate. Unfortunately, this leads to the emergence of closed timelike curves in the spacetime. Thus we will interpret the singularity following \cite{Bonnor:NUT} and shall ascribe the conical singularity to a spinning rod. This explicitly demonstrates that for $\ell>a$ one removes the curvature singularity, but the conical singularity remains. Thus the solution though devoid of curvature singularity cannot be interpreted as a regular rotating black hole solution. Note that it is the NUT parameter that abhors curvature singularity in general, it gets tamed when rotation dominates over it, i.e, when $a \geq l$. For $\ell=0$, we get back the ring singularity of Kerr (or, Kerr-Newman) located at $r=0$ and $\theta=\pi/2$. 

Curiously, there exists one subclass of the solution for which the location of the event horizon coincides with that of the Schwarzschild, located at $r=2M$. As evident from the expression for $\Delta$ it is clear that this will happen when $Q^{2}=l^{2}-a^{2}$, which in turn demands $l \geq a$. Thus whenever the NUT charge is larger than (or, equal to) the black hole rotation parameter, a suitable choice for the electric charge will lead to such a configuration for which either there is no curvature singularity for $l>a$ or if it exists (for $l=a$), it would be spacelike. This is a very novel and interesting aspect that merits consideration of its own accord. Out of the four parameters $(M,l,a,Q)$, the former two are gravitational charge and are gravitationally attractive while the latter two are non-gravitational and repulsive. For the prescription $Q^2 + a^2 = l^2$, they perfectly and effectively balance each other leading to either a black hole with spacelike singularity and horizon located at $r=2M$ or to a black hole solution free of curvature singularity. Both of these are special cases in the Kerr-Newman-NUT family. Note that out of the four hairs, $M$ and $Q$ only enter in $\Delta$, which characterizes the coloumbic aspect, while in contrast, $l$ and $a$ in addition to $\Delta$ also participate in defining the spacetime symmetry and henceforth construct the magnetic aspect. Given its structural simplicity and special location of the event horizon, it will be very interesting to understand various intriguing features this spacetime has to offer. Let us now specialize the metric presented in \ref{metric_KN_old} with $Q^2=l^2-a^2$, which leads to,
\begin{align}\label{metric_KN}
ds^{2}=-\frac{\Delta}{\rho ^{2}}\left(dt-Pd\phi\right)^{2}
+\frac{\sin ^{2}\theta}{\rho ^{2}}\left\{\left(r^{2}+a^{2}+l^{2}\right)d\phi -adt\right\}^{2}
+\frac{\rho ^{2}}{\Delta}dr^{2}+\rho ^{2}d\theta ^{2}
\end{align}
with $\Delta = r^{2}-2Mr$. Though the horizon is only determined by the mass $M$, and coincides with that of the Schwarzschild's, yet the black hole has non-zero rotation parameter as well as inherits both electric and NUT charges. As \ref{metric_KN} suggests, the above black hole geometry can be described by essentially three hairs, namely, the black hole mass $M$, rotation parameter $a$ and the \N\ parameter $l$. However there is indeed an electric charge $Q^{2}=l^{2}-a^{2}$, but is not manifest in the metric structure. The duality between the gravitational mass $M$ and the gravitomagnetic mass $l$ will exist in this spacetime as well. The line element is still given by \ref{metric_KN}, with only $\Delta$ being modified to $\Delta=r(r-2M)$. As emphasized earlier, this is in stark contrast to a generic Kerr-Newman-NUT spacetime, since in this case there is a single black hole horizon located at $r=2M$, without any possibility of having naked singularity for any values of $a$ or $l$ whatsoever. Furthermore, the singularity associated with the spacetime only exists for $l=a$ at $r=0,\theta=\pi$ and is spacelike unlike the situation with general \KNN\ spacetimes. Surprisingly, the curvature singularity does not exist for $l>a$, as the equation $\cos \theta=-l/a$ does not have any solution for $l>a$. Hence the above spacetime with $Q^{2}=l^{2}-a^{2}$ with $l>a$ represents a black hole solution, with an event horizon, but without any curvature singularity. As expected, the above spacetime turns out to be an exact solution of the Einstein-Maxwell field equations, with the Maxwell field tensor having the following form,
\begin{align}
F&=\sqrt{l^{2}-a^{2}}\Bigg\{\dfrac{1}{\rho^4} \Big\{r^2+l^2-a^2 \cos^2\theta\Big \}\Big(1+\dfrac{4a l \cos\theta \rho^2}{(r^2+l^2-a^2 \cos^2\theta)^2}\Big)^{1/2} dr \wedge \Big[dt-a \sin^2\theta d\phi\Big]
\nonumber 
\\
&+\dfrac{2ar\sin\theta \cos\theta}{\rho^4} d\theta \wedge \Big[(r^2+a^2+l^2)d\phi-a dt\Big]\Bigg\}.
\end{align}
where, \enquote*{$\wedge$} is defined as the outer product. This suggests several interesting points regarding this spacetime. First of all, the \N\ parameter cannot be set to zero, because $l^{2}\geq a^{2}$, without setting $a=0$, resulting into Schwarzschild spacetime. On the other hand when $a=0$, we end up with the Reisnner-Nordstr\"{o}m-NUT black hole.

\begin{figure}[htp]
\centering
\fbox{\includegraphics[scale=.9]{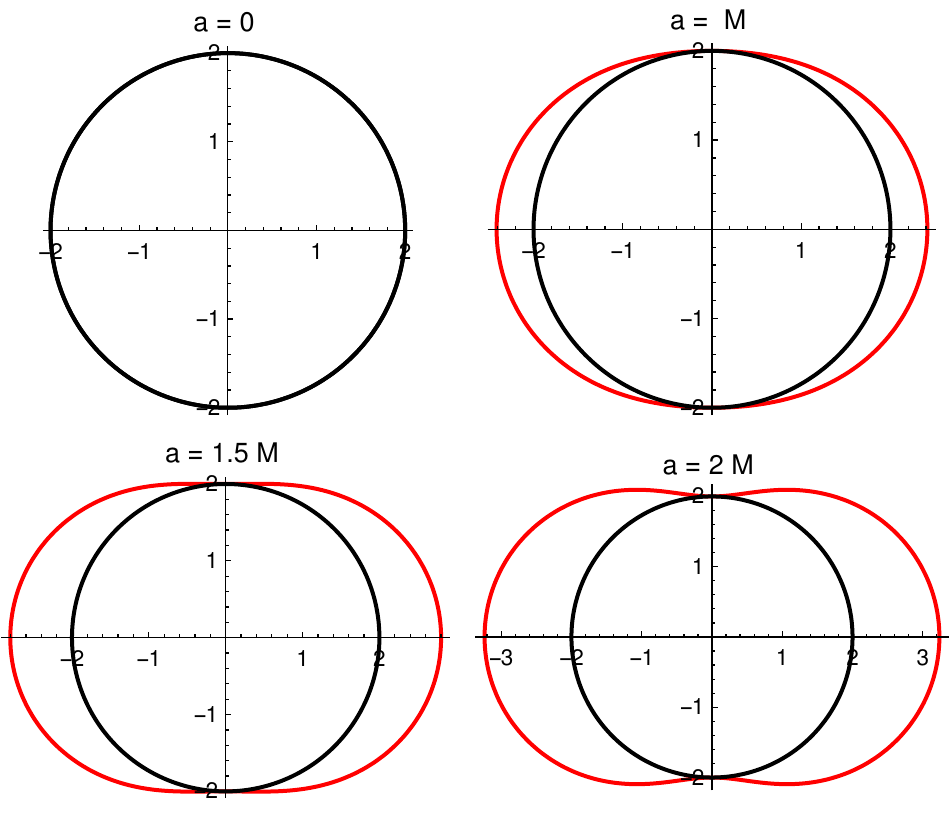}}
\caption{The event horizon (black curve) and the \ER~is shown (red outer curve) in a \KNN~spacetime. Event horizon is always located at $r_{\rm eh}=2M$ and it is independent of the angular momentum as well as \N~parameter. The \ER~is a function of \enquote*{a} and independent of the \N~parameter \enquote*{$l$}. }
\label{Fig_01}
\end{figure}

Due to presence of the rotation parameter it follows that, the spacetime consists of an \ER\ alongside the usual event horizon at $r=2M$. The boundary surface of the ergoregion can be obtained by solving the equation $g_{tt}=0$, which in this case translates into $\Delta-a^2 \sin^2\theta=0$. This being a quadratic equation in the radial coordinate $r$, yields two solutions related to the outer and the inner ergoregion boundaries respectively,
\begin{equation}
r_{\rm ergo-outer}=M+\sqrt{M^2+a^2 \sin^2\theta}, \qquad \text{and} \qquad r_{\rm ergo-inner}=M-\sqrt{M^2+a^2 \sin^2\theta}.
\end{equation}
It is easy to see that the inner boundary is unphysical as it is situated at $r<0$ and hence we will only refer to the outer boundary henceforth. The location of the ergoregion and the horizon are presented in \ref{Fig_01}. As evident, for $a=0$, the ergoregion ceases to exist and the horizon is located at $r=2M$. With the increase in the angular momentum of the black hole, the ergoregion starts to surround the event horizon. The width of the ergoregion increases with the increase of angular momentum. Note that neither the event horizon nor the ergoregion depends on the value of the \N\ charge $l$ because of the following relation: $Q^{2}+a^{2}=l^{2}$. In this case, the ergoregion is bounded as $2M\leq r \leq M(1+\sqrt{1+a^2/M^2})$ while for the Kerr, it is $M(1+\sqrt{1-a^2/M^2}) \leq r \leq 2M$. That is, the lower limit for the ergoregion of a \KNN\ black hole with $l=a$ is the upper limit for the ergoregion of a Kerr black hole.

\section{Trajectory of massive and massless particles}
\label{KNN_Trajectory}

Understanding the geometry of any spacetime requires a thorough analysis of the trajectory of a massive as well as massless particles. Keeping this in mind, in this section we shall derive the basic equations describing the trajectory of a test particle in the \KNN~spacetime presented above. In the spirit of Kerr geometry, the geodesic motion is completely integrable in a \KNN~spacetime as well due to the presence of the Carter constant \cite{Carter:1968rr}. This can be easily shown by investigating the Hamilton-Jacobi equation and hence establishing the separability of the radial and the angular part. Due to invariance of the metric elements under time translation and rotation with $\phi$ as the rotation angle, we can treat the energy $E$ and the angular momentum $L$ as conserved quantities. This demands to write down the action associated with the motion of the particle in the following manner,
\begin{align}
\mathcal{A}=-Et+L\phi +\mathcal{A}_{r}(r)+\mathcal{A}_{\theta}(\theta)
\end{align}
Here $\mathcal{A}_{r}(r)$ and $\mathcal{A}_{\theta}(\theta)$ correspond to the parts of the action dependent on radial coordinate $r$ and angular coordinate $\theta$ respectively. Given the above structure of the action associated with a particle moving in the \KNN\ spacetime, the Hamilton-Jacobi equation becomes separable. Thus one can obtain separate equations for $\mathcal{A}_{r}(r)$ and $\mathcal{A}_{\theta}(\theta)$ respectively, having the following structures,
\begin{align}
\left(\frac{d\mathcal{A}_{\theta}}{d\theta}\right)^{2}&+\frac{(EP-L)^{2}}{\sin ^{2}\theta}+m^{2}(l+a\cos \theta)^{2}=K
\label{KNN_Tra_01}
\\
\Delta \left(\frac{d\mathcal{A}_{r}}{dr}\right)^{2}&+m^{2}r^{2}-\frac{1}{\Delta}
\left\{E(r^{2}+a^{2}+l^{2})-aL\right\}^{2}=-K
\label{KNN_Tra_02}
\end{align}
In the above expressions the quantity $K$ acts as the separation constant and $m$ is the mass of the orbiting particle. Among others $l$ is the NUT charge, $a$ is the rotation parameter and $P=a\sin ^{2}\theta-2l\cos \theta$. Since the components of the momentum four vector of the orbiting particle correspond to, $\partial \mathcal{A}/\partial x^{\mu}$, it is possible to rewrite \ref{KNN_Tra_01} and \ref{KNN_Tra_02} in a more explicit form. This involves writing the separation constant $K$ appearing in both \ref{KNN_Tra_01} and \ref{KNN_Tra_02}, such that $K=\lambda +(L-aE)^{2}$, with $\lambda$ as the Carter constant. Thus we obtain the following geodesic equations,
\begin{align}
m^{2}{\rho ^{4}}\left(\frac{dr}{d\tau}\right)^{2}
&=\left\{E(r^{2}+a^{2}+l^{2})-aL\right\}^{2}-\left(\lambda +m^{2}r^{2}\right)\Delta -\Delta \left(L-aE\right)^{2} 
\label{EOM_Radial}
\\
m^{2}{\rho ^{4}}\sin^{2}\theta \left(\frac{d\theta}{d\tau}\right)^{2}
&=\lambda \sin ^{2}\theta+\left(L-aE\right)^{2}\sin ^{2}\theta -\left[EP-L\right]^{2}-m^{2}\sin ^{2}\theta (l+a\cos \theta)^{2} 
\label{EOM_Angular}
\end{align}
where $\tau$ represents the affine parameter for a timelike geodesic. Similar to the above case with massive particles, for photons with zero rest mass, the geodesic equations become,
\begin{align}
{\rho ^{4}}\left(\frac{dr}{d\nu}\right)^{2}
&=\left\{E(r^{2}+a^{2}+l^{2})-aL\right\}^{2}-\lambda \Delta -\Delta \left(L-aE\right)^{2} 
\label{EOM_Radial_photon}
\\
{\rho ^{4}}\sin^{2}\theta \left(\frac{d\theta}{d\nu}\right)^{2}
&=\lambda \sin ^{2}\theta+\left(L-aE\right)^{2}\sin ^{2}\theta -\left[EP-L\right]^{2}
\label{EOM_Angular_photon}
\end{align}
where $\nu$ stands for affine parameter along the null geodesic. Given the above two geodesic equations one can proceed towards solving these equations, which will ultimately lead to the trajectory of a particle moving in the \KNN\ spacetime. With the complicated structure of the geodesic equations as presented above, it is very difficult to solve them in general circumstances. However in some specific situations, e.g., in the equatorial plane (located at $\theta =\pi/2$) it is indeed possible to solve the above equations analytically, which we will discuss next. Having grasped the analytical structure of the trajectory associated with the equatorial plane, we will consider the general scenario later on. 
\subsection{Orbits confined on a given plane}
\label{KNN_Tra_Equatorial}
The addition of the NUT charge will introduce nontrivial difficulties while obtaining the orbital dynamics for a massive or even a massless particle. This can be understood by employing the angular equations given in \ref{EOM_Angular} and \ref{EOM_Angular_photon} respectively. For any trajectory to be confined on a particular plane $\theta=\theta_{0}$, we bound to have $\dot{\theta}=\ddot{\theta}=0$. While the condition $\dot{\theta}=0$ indicates a trajectory moving on a constant plane, the additional second derivative ensures that the first derivative remains null throughout its motion. Therefore both of these conditions are essentials to determine any planner orbit in the presence of a NUT charge and unlike the Kerr black hole, this constraint would introduce stringent bound on the particle trajectories in \KNN~spacetime. In the upcoming discussions, we shall explore any possible scenarios in which the conditions for a planner orbit can be satisfied in a certain ranges of parameters. Regarding that, we first introduce the massless particles and following that, the phenomenon involving massive particles will be addressed.

\noindent
\subsubsection{The massless particles}
\label{KNN_Tra_Eqa_Massless}

For a massless particle, the effective radial potential can be written as
\begin{align}
\mathcal{V}_{\rm eff}(r)=\left\{E(r^{2}+a^{2}+l^{2})-aL\right\}^{2}-\Delta \left\{\lambda+ \left(L-aE\right)^{2} \right\}.
\label{KNN_Equ_massl_eff}
\end{align}
Remember, the value of the Carter constant $\lambda$ would only be fixed from the angular equations, i.e,  $\dot{\theta}=\ddot{\theta}=0$. Even though for a general $E$ and $L$, the angular equations are complicated to solve and often subjected to numerical calculations, we  start with a simple exercise which corresponds to the following relation between angular momentum and energy: $L=aE$. In this case the impact parameter, defined as the ratio of angular momentum and energy, turns out to be $D_{\rm sp}\equiv L/E=a$. On other hand, the angular equations become 
\begin{eqnarray}
\mathcal{V}_{\rm eff}(\theta) &=& \lambda \sin^2\theta-\left\{aE \sin^2\theta-2 E l \cos\theta-aE\right\}^2 = 0, \nonumber \\
\dfrac{\mathcal{V}_{\rm eff}(\theta)}{d\theta} &=& 2 \sin\theta \cos\theta \lambda -2 \left\{aE \sin^2\theta-2 E l \cos\theta-aE\right\}\left\{2 a E \sin\theta \cos\theta+2 E l \sin\theta\right\}=0. 
\label{eq:angular_constraint}
\end{eqnarray}
It can be easily noticed that the above equations has an obvious solution with $\lambda=0$ and $\theta=\pi/2$ and therefore, the particle with an impact parameter equal to the rotation parameter of the black hole may exist on the equatorial plane. By setting $\lambda=0$, one can establish the following results regarding the trajectory of the photon,
\begin{align}
\dot{r}&= \pm~ E, \qquad \dot{t}=\dfrac{E(r^2+a^2+l^2)}{\Delta}, \qquad \dot{\phi}=\dfrac{a E}{\Delta}~,
\end{align}
where \enquote*{dot} defines derivative with respect to some affine parameter $\nu$ associated with null geodesics. Note that as the rotation parameter $a$ vanishes, the angular momentum also goes to zero. Thus in the case of vanishing rotation parameter, the photon with the above impact parameter moves along radial geodesics. Hence by analogy, the null geodesics in the present context as well are dubbed as \emph{radial-like} geodesics. Further, the $\pm$ sign in the expression for $\dot{r}$ denotes both inward as well as outward motion. Having dealt with this special case, we now concentrate on a more important aspect of these null geodesics, namely the location of circular photon orbits. 
\paragraph{Circular photon orbits :}

Let us now describe the structure of photon circular orbits in the \KNN~spacetime with $\Delta=r^2-2Mr$. The necessary and sufficient conditions for the existence of a circular orbit are given by,
\begin{equation}
\dot{r}=\mathcal{V}_{\rm eff}(r)=0, \qquad \text{and} \qquad \ddot{r}=\dfrac{1}{2}\dfrac{d\mathcal{V}_{\rm eff}(r)}{dr}=0~.
\label{Condition_Circular}
\end{equation} 
Here $\mathcal{V}_{\rm eff}(r)$ corresponds to the effective potential presented in \ref{KNN_Equ_massl_eff}. In passing, it should be carefully noted that any solution to the above equation has to be compatible with the angular constraint given in \ref{eq:angular_constraint}.

The presence of a nonzero Carter constant will forbid to write the above equations as a function of the impact parameter $D_{\rm ph}=L/E$ and therefore we seek for general expressions for both energy and angular momentum. We start by writing $L=aE+x$ and employ the circular orbit conditions as
\begin{eqnarray}
\mathcal{V}_{\rm eff}(r)=\left\{E(r^2+l^2)-ax \right\}^2-\Delta( x^2+\lambda)=0, \quad \text{and} \quad \mathcal{V}^{\prime}_{\rm eff}(r)=\dfrac{d\mathcal{V}_{\rm eff}(r)}{dr}= 4Er\bigl\{E(r^2+l^2)-ax\bigr\}\\ \nonumber
\qquad -2(r-M)(x^2+\lambda)=0. 
\end{eqnarray}
From the above set of equations, the expression for energy can be calculated by computing the expression $2r\mathcal{V}_{\rm eff}(r)-(r^2+l^2)\mathcal{V}^{\prime}_{\rm eff}(r)=0$
\begin{equation}
E^2=\dfrac{1}{(r^2+l^2)^2}\left\{a^2 x^2+M r( x^2+\lambda)+l^2 (1-M/r)(x^2+\lambda) \right\},
\label{eq:express_energy}
\end{equation}
and further substituting the above into the equation $4r\mathcal{V}_{\rm eff}(r)-(r^2+l^2)\mathcal{V}^{\prime}_{\rm eff}(r)=0$, we arrive at
\begin{eqnarray}
&  (x^2+\lambda)\Bigl\{l^4(r-M)^2(x^2+\lambda)-2l^2 r^2(x^2+\lambda)(3M^2-4 M r+r^2)+ \nonumber \\
&  \qquad \qquad \qquad \qquad \qquad \qquad \qquad r^4(r-3M)^2(x^2+\lambda)+4 a^2 x^2 r^3 (2M-r)\Bigr\}=0.
\end{eqnarray}
This is the final equation dictating the location of the circular orbits in the presence of a NUT charge. From above, we can have either of the possibilities
\begin{eqnarray}
x^2+\lambda=0,\quad \text{and} \quad (\lambda+x^2)\Bigl\{l^4(r-M)^2-2 l^2 r^2 (r^2-4 M r+3 M^2)+r^4 (r-3M)^2\Bigr\}+\nonumber \\
 4 a^2 x^2 r^3 (2 M-r)=0.
 \label{eq:options}
\end{eqnarray}
\noindent
In the first case with $x^2+\lambda=0$, the expression for energy and angular momentum become
\begin{equation}
E=\pm\dfrac{ax}{r^2+l^2},\quad  \text{and} \quad L=\dfrac{x(r^2+a^2+l^2)}{r^2+l^2},
\label{eq:energy_momentum}
\end{equation}
with $x=\pm \sqrt{|\lambda|}$. On the other hand, both the angular equations, i.e, $\dot{\theta}=\ddot{\theta}=0$ will be satisfied for the following equality
\begin{equation}
aE \sin^2\theta-2lE\cos\theta-L=0,
\end{equation}
and as it can be easily supplemented with the expressions given in \ref{eq:energy_momentum}, we arrive at either of these possibilities
\begin{eqnarray}
aE \sin^2\theta-2lE\cos\theta-L=-\dfrac{x\rho^2 }{r^2+l^2}=0, \quad \text{for}~ E = \dfrac{ax}{r^2+l^2}, \nonumber \\
2(r^2+a^2+l^2)-\rho^2=r^2+a^2 \sin^2\theta+(l-a)^2=0, \quad \text{for}~E = \dfrac{-ax}{r^2+l^2},
\end{eqnarray}
From the above, it is easy to notice that neither of these cases  would lead to a consistent outcome. Therefore, it is unexpected to encounter any circular photon orbit which obeys the condition $x^2+\lambda=0$.


In the second case given in \ref{eq:options}, we have the following expressions for $x$,
\begin{eqnarray}
x^2 &=& \dfrac{\lambda\left\{r^2(r-3M)+l^2(M-r)\right\}^2}{2 l^2 r^2 (r^2+3 M^2-4 M r)-r^3 \Big[a^2 (8M-4r)+r(r-3M)^2 \Big]-l^4(r-M)^2},
\end{eqnarray}
and as it can be easily noticed that both \enquote*{$+$} and \enquote*{$-$} values of $\lambda$ are possible with appropriate signature of the denominator. With the above expression, we shall now be able to express the energy (using \ref{eq:express_energy}) and momentum in terms of radial distance and black hole's parameters. 
\begin{eqnarray}
E^2 &=& \Bigg(\dfrac{a^2(r-M)^2\lambda}{2 l^2 r^2 (r^2+3 M^2-4 M r)-r^3 \Big[a^2 (8M-4r)+r(r-3M)^2 \Big]-l^4(r-M)^2}\Bigg), \nonumber \\
L &=& \pm \dfrac{\lambda^{1/2}  \left\{(a^2 \mp l^2)(M-r)+r^2(r-3M)\right\}}{\Big\{2 l^2 r^2 (r^2-4 M r+3 M^2)-l^4(r-M)^2-r^3\Bigl[4 a^2 (2M-r)+r(r-3M)^2\Bigr]\Big\}^{1/2}}. \nonumber \\
\label{eq:energy_momentum_02}
\end{eqnarray}
The key point to consider is that both the expressions are proportional to the Carter constant $\lambda$ and while obtaining the angular equations, it is expected that $\lambda$ would not have any active effect. In addition, for the expression of energy, we only consider the \enquote*{+} sign as that would contribute a positive energy for circular orbits located at $r>r_{\rm eh}=2M$.

The constraint from the angular potential will lead to the following expression,
\begin{eqnarray}
\mathcal{V}_{\rm eff}(\theta)=\lambda \mathcal{Q}^{-1} \Bigg[\mathcal{Q}\sin^2\theta+ \Big\{l^2(M-r)+r^2(r-3M)+a^2(r-M)+a(M-r)(a\sin^2\theta-2l \cos\theta)\Big\}^2\nonumber \\
-\Big\{l^2(M-r)+r^2(r-3M)\Big\}^2 \sin^2\theta\Bigg]=0,
\end{eqnarray}
and $\dfrac{d\mathcal{V}_{\rm eff}}{d(\theta)}=0$ in which, $\mathcal{Q}=l^4(r-M)^2-2l^2 r^2 (r^2-4M r+3M^2)+r^4(r-3M)^2+a^2 r^3(8M-4r)$. It should be mentioned again that the angular part is completely independent of the  Carter constant and therefore, both \enquote*{+} and \enquote*{$-$} values $\lambda$ can be relevant depending on the appropriate sign of the denominator in \ref{eq:energy_momentum_02}. For example with $l=a=M$, the approximate radius and inclination of one of the circular orbits become $r_{\rm c}=4.7147M$, $\theta_{\rm c}=100^{\circ}$ respectively. Substituting these values of $r_c$ and $\theta_c$, we obtain the conserved quantities as $\Big\{E,L\Big\}=\Big\{0.55 (-\lambda M^{-2})^{1/2},5.66 M (-\lambda M^{-2})^{1/2}\Big\}$ and needless to say $\lambda$ has to be negative.
\subsubsection{The timelike geodesics}
\label{KNN_Equ_time}
Having demonstrated the photon circular orbit in the context of massless particles, let us take up the case for massive particles. Similar to the previous occasion with massless particles, we follow the identical technique to describe the circular orbits in the \KNN~black hole. We start with the effective radial potential given as, 

\begin{align}
V_{\rm eff}(r)=\left[\tilde{E}(r^{2}+a^{2}+l^{2})-a\tilde{L}\right]^{2}-\Delta \left(\tilde{L}-a\tilde{E}\right)^{2}-\Delta (r^{2}+\lambda).
\end{align}
and the $\lambda$ as usual will be fixed by the angular equations $\dot{\theta}=\ddot{\theta}=0$, here the \enquote*{dot} defines the derivative with respect to the proper time $\tau$. However, before obtaining the circular orbits for a general energy and momentum, we consider a special case such that, $\tilde{L}=a\tilde{E}$ and we arrive at
\begin{equation}
\dot{r}=\pm~ \sqrt{\tilde{E}^2-\dfrac{\Delta (r^2+\lambda)}{(r^2+l^2)^2}}; \quad \dot{t}=\dfrac{\tilde{E}(r^2+a^2+l^2)}{\Delta}; \qquad \dot{\phi}=\dfrac{a \tilde{E}}{\Delta}.
\end{equation}
As the rotation parameter vanishes, the angular velocity will vanish as well, which in turn depicts the motion along radial geodesics. Following this analogy, the above trajectory of massive particles depict \emph{radial-like} geodesics. Further note that the expression for $\dot{r}$ associated with the geodesic of a massive particle  differs from the one for massless particles, by the factor proportional to $\Delta$. Thus on the black hole horizon, $\Delta$ vanishes and hence the geodesic becomes null as expected. 

On the other hand, the angular potential would take the form
\begin{equation}
V_{\rm eff}(\theta)=\tilde{\lambda} \sin^2\theta-\tilde{E}^2 \left\{a \sin^2\theta-2l \cos\theta-a\right\}^2-\sin^2\theta (l+a \cos\theta)^2,
\end{equation} 
and for a motion confined on a given plane, we require to have
\begin{equation}
V_{\rm eff}(\theta)=\dfrac{dV_{\rm eff}(\theta)}{d\theta}=0.
\end{equation}
\noindent
For a given value of $a=l=M$, we obtain that the marginally bound massive geodesics with $\tilde{E}=1$ can be confined on a plane given as $(\theta_c,\lambda_c) \approx (104^{\circ},0.77M^2)$.

Having described this particular case in which the impact parameter is directly related to black hole spin, let us now concentrate on the motion of massive particles moving in circular orbits. 
\paragraph{Circular orbits for timelike particles:}

\begin{figure}[htp!]
\subfloat[An appropriately scaled Effective potential for a timelike geodesic with $\tilde{E}=1$ and $\tilde{L}=3.6M$.\label{Fig_l_2a}]{\includegraphics[height=6cm,width=.48\linewidth]{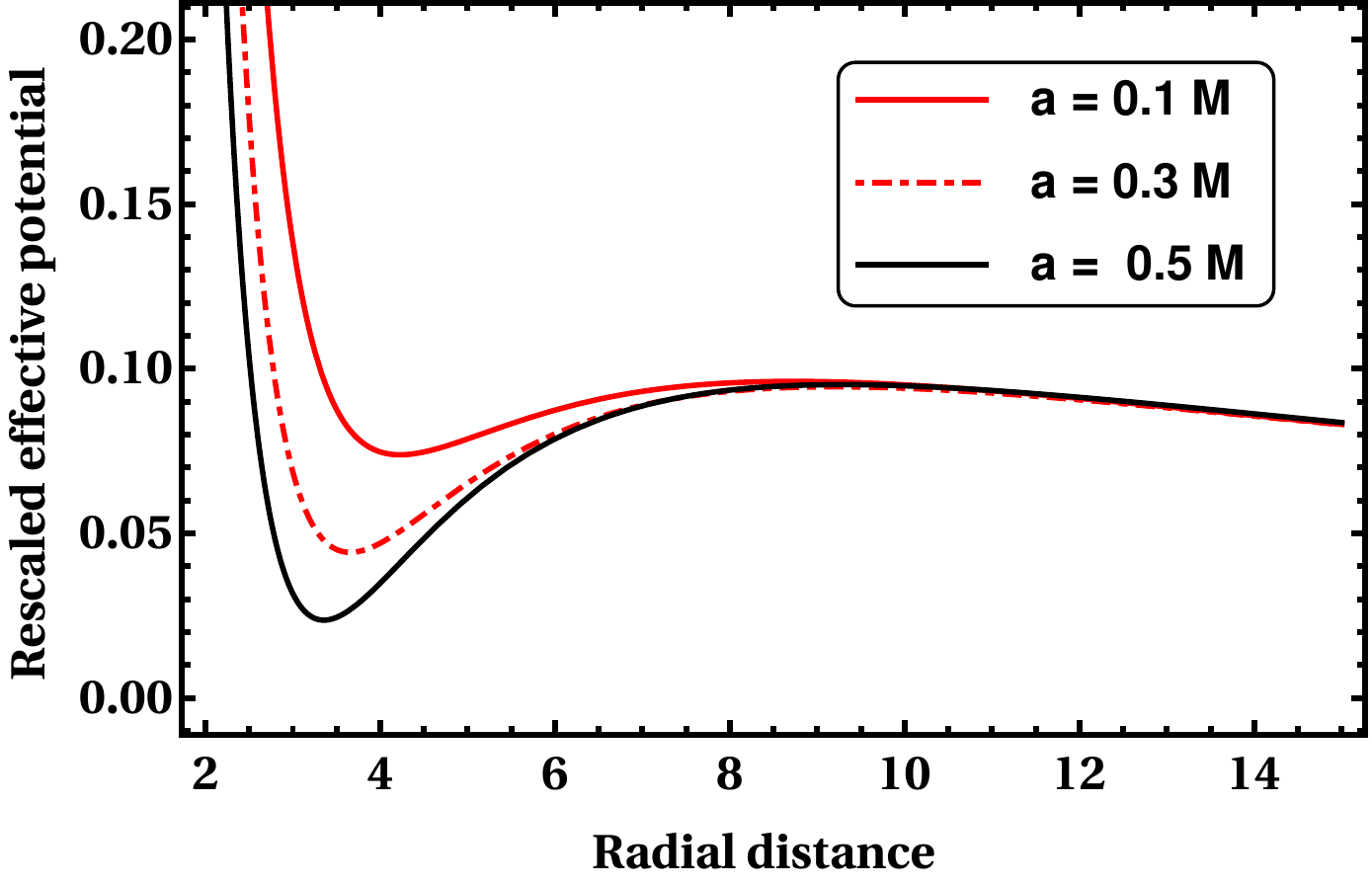}}
\hfill
\subfloat[The rescaled effective potential for timelike geodesics has been presented with $\tilde{E}=1$ and $\tilde{L}=3.7M$.\label{Fig_l_a}]{\includegraphics[height=6cm,width=.48\linewidth]{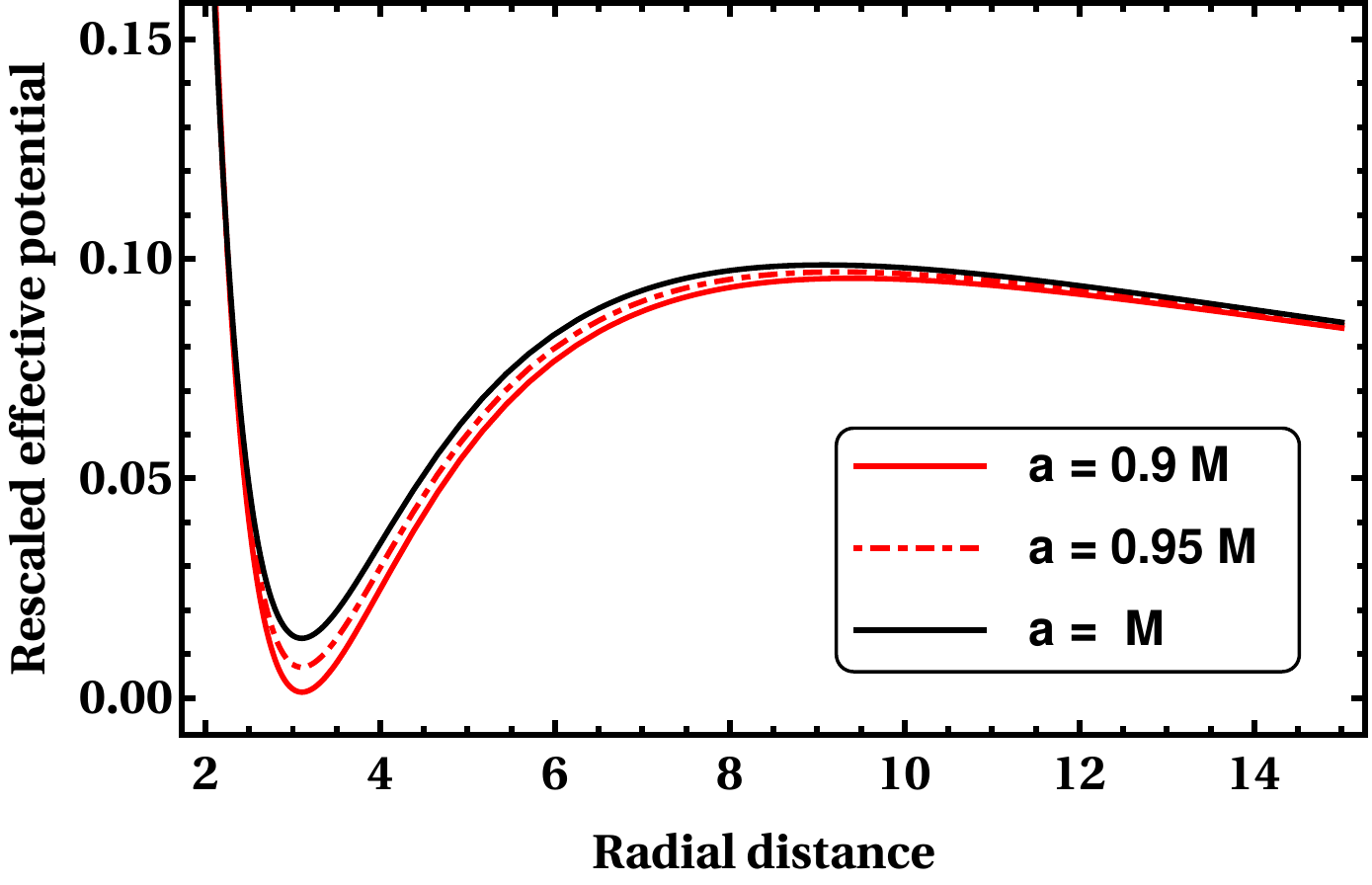}}
\caption{The effective potential ($V_{\rm eff}$), rescaled using $(r^{2}+l^{2})^{-2}$, is presented in the above figure for marginally bound orbits (i.e., $\tilde{E}=1$) and various choices of angular momentum of the black hole. We discuss two particular cases of interest, one is $\tilde{L}=3.6M$ and another corresponds to $\tilde{L}=3.7M$. In both these plots, the NUT charge, Carter constant and black hole rotation parameters are taken to be identical. The plots explicitly depict presence of circular orbits, given the existence of minima of the potential.}
\label{Fig_Potential}
\end{figure}

In this section, we shall discuss the possible existence of timelike circular orbits and hence compute the exact expressions for conserved quantities associated with it. In this case as well, the conditions for circular orbit are given by $V_{\rm eff}=0$ and $V_{\rm eff}'=0$, see \ref{Condition_Circular}. To see the existence of circular orbits in this spacetime in a direct manner, we have plotted the effective potential $V_{\rm eff}$, rescaled by $(r^{2}+l^{2})^{-2}$, for specific choices of various constants of motion and parameters appearing in the problem in \ref{Fig_Potential}. In particular, we have considered only marginally bound orbits, i.e., orbits with $\tilde{E}=1$ and the situation when the rotation parameter and NUT charge of the black hole coincides, implying $l=a$ and $\tilde{\lambda}=M^2$. As evident from \ref{Fig_Potential} the effective potential indeed exhibits a minima for various choices of the $(a/M)$ as well as $(\tilde{L}/M)$ ratio and hence allows existence of stable circular orbits in the spacetime. Thus we have explicitly demonstrated the existence of stable circular orbits in this spacetime, which we will now explore analytically.

In order to determine the location of the circular orbits, we need to solve for the conditions $V_{\rm eff}=0$ and $V_{\rm eff}'=0$ analytically. This can be achieved by substituting $\tilde{L}=a\tilde{E}+x$ and hence computing $x$ from the circular orbit conditions as given above, resulting into \cite{chandrasekhar1998mathematical},
\begin{align}
V_{\rm eff}(r)&=\tilde{E}^2\left(r^2 + l^2\right)^{2}+2M r^{3}-r^{4}-2a\tilde{E} \left(r^{2}+l^{2}\right)x+a^{2}x^{2}+2Mrx^{2}-r^{2}x^{2}-\tilde{\lambda}\Delta=0
\nonumber
\\
V_{\rm eff}'(r)&=4\tilde{E}^{2}r \left(l^2 + r^2\right)+6Mr^{2}-4r^{3}-4a\tilde{E}rx+2Mx^{2}-2rx^{2} -\tilde{\lambda} \left(2M-2r\right)=0. 
\end{align}
Here, the `prime' denotes a derivative with respect to the radial coordinate and $\Delta=r^2-2Mr$. Given the above equations one can easily solve for the energy $\tilde{E}$ by recombining both the above equations such that the quadratic terms cancel away. This can be achieved by considering the following combination: $4rV_{\rm eff}(r) -(r^2+l^2)V_{\rm eff}'(r)=0$. This will yield,
\begin{equation}
\tilde{E}=\Big\{l^2 \Big[r(2r^2+x^2+\lambda)-M(3 r^2+x^2+\lambda)\Big]+r\Big[2 a^2 x^2-r^2 (x^2+\lambda)+M r (r^2+3 x^2+3 \lambda)\Big]\Big\}
\left\{2 a r (l^2 + r^2) x\right\}^{-1}
\label{Energy_timelike}
\end{equation} 
However the above expression for energy involves the unknown quantity $x$, determination of which is essential for an estimation of angular momentum as well. This can be achieved by substituting the expression for energy as in \ref{Energy_timelike} into the equation, $2rV_{\rm eff}(r)-\left(r^2+l^2\right)V_{\rm eff}'(r)=0$. This results into the following quartic equation for $x$, which reads,
\begin{equation}
A x^4+B x^2+C=0.
\label{Quadratic}
\end{equation}
The coefficients appearing in the above expression are functions of the radial distance as well as mass angular momentum and NUT charge of the black hole. Introducing $u=1/r$, we finally obtain, 
\begin{align}
\dfrac{A}{u^2}&=M^2 u^2\left(l^2 u^2-3\right)^2+\left\{u^2l^2-1\right\}^2 
-4la^2u^2-2Mu\Big[3+u^2\Big\{-4a^2+l^2\left(-4+u^2l^2\right)\Big\}\Big]
=Z_{+}Z_{-}
\nonumber 
\\
\dfrac{B}{2u}&=u \Big[-2 a^2 (1+u^2 \lambda)+(1-l^2 u^2)(\lambda-2 l^2-\lambda u^2 l^2)\Big]+M^2 u (3-l^2 u^2) \Big[1-u^2(3l^2-3 \lambda+\lambda u^2 l^2)\Big]\nonumber \\
& ~~~~~ + M \Big[-1+u^2 (4 a^2+4 a^2 u^2 \lambda-6\lambda)+10 l^2+u^2 l^2 (8 \lambda-5 l^2-2 \lambda u^2 l^2)\Big],
\nonumber 
\\
C &=\Big[u\left\{\lambda-2 l^2-u^2 l^2 \lambda \right\}+M \left\{-1-3 u^2\lambda+3 u^2 l^2+l^2 u^4 \lambda\right\} \Big]^2.
\end{align}
In the above expression the quantity $A$ can be written as a product of two quantities denoted as $Z_{\pm}$ with the following expressions for each of them,
\begin{equation}
Z_{\pm}=(1+l^2 u^2)^{-1}\Big[\left\{1-l^4 u^4 +M u(l^4 u^4-2 l^2 u^2-3)-2 a^2 u^2 (1+\lambda u^2)\right\}\pm 2au^{3/2}\mathcal{K}^{1/2}\Big],
\end{equation} 
where, 
\begin{equation}
\mathcal{K}=M(1+l^2 u^2)\left\{1+3 \lambda u^2-3 l^2 u^2-\lambda l^2 u^4\right\}+u \left\{2 l^2-\lambda+l^4 u^2(2+u^2 \lambda)+a^2 (1+\lambda u^2)^2\right\},
\end{equation}
Using the above expressions for the three quantities, namely $A$, $B$ and $C$ it is certainly possible to determine a solution for $x$ given \ref{Quadratic}. The solution turns out to be much simpler by defining $\mathcal{G}_{u}=1-2M u$, resulting into,
\begin{equation}
x^2 u^2 =\dfrac{1}{Z_{+}Z_{-}}\Big \{(1+l^2u^2)\mathcal{G}_{u} Z_{\pm}-Z_{+}Z_{-}(1+\lambda u^2) \Big \}.
\end{equation}
It turns out that the term within square bracket can be written in a nice form, such that the solution for $x$ itself can be presented in a simplified manner as,
\begin{equation}
x_{\pm}=\mp\dfrac{1}{\sqrt{u Z_{\pm}}}\dfrac{1+\lambda u^2}{(1+l^2 u^2)^{1/2}}\left\{a\sqrt{u}\mp (1+\lambda u^2)^{-1}\mathcal{K}^{1/2}\right\}.
\end{equation}
Therefore we have succeeded in determining the quantity $x$ and hence the energy and angular momentum associated with the circular motion can be easily obtained. In particular the expression for energy of a particle moving in a circular orbit of radius $r=r_{c}$ takes the following form,
\begin{equation}
\tilde{E}_{c}^{\pm}=(1+l^2 u^2_{\rm c})^{-1}\Big[1+l^2 u^2_{\rm c}-2 M u_{\rm c} (1+l^2 u^2_{\rm c})-a^2 u^2_{\rm c}(1+\lambda u^2_{\rm c})\pm   a u^{3/2}_{\rm c}\mathcal{K}^{1/2}\Big] \left\{Z_{\pm}^{c}(1+l^2u_{c}^2)\right\}^{-1/2}.
\label{energy_circular}
\end{equation}
Here all the quantities have been evaluated at the circular orbit radius $r=r_{c}$. The two signs present in the above expression denotes energy for direct and retrograde orbits respectively. 

\begin{figure}[htp]
\subfloat[The figure shows the co-rotating circular orbits for $l=a$. \label{Fig_7a}]{\includegraphics[height=5cm,width=.49\linewidth]{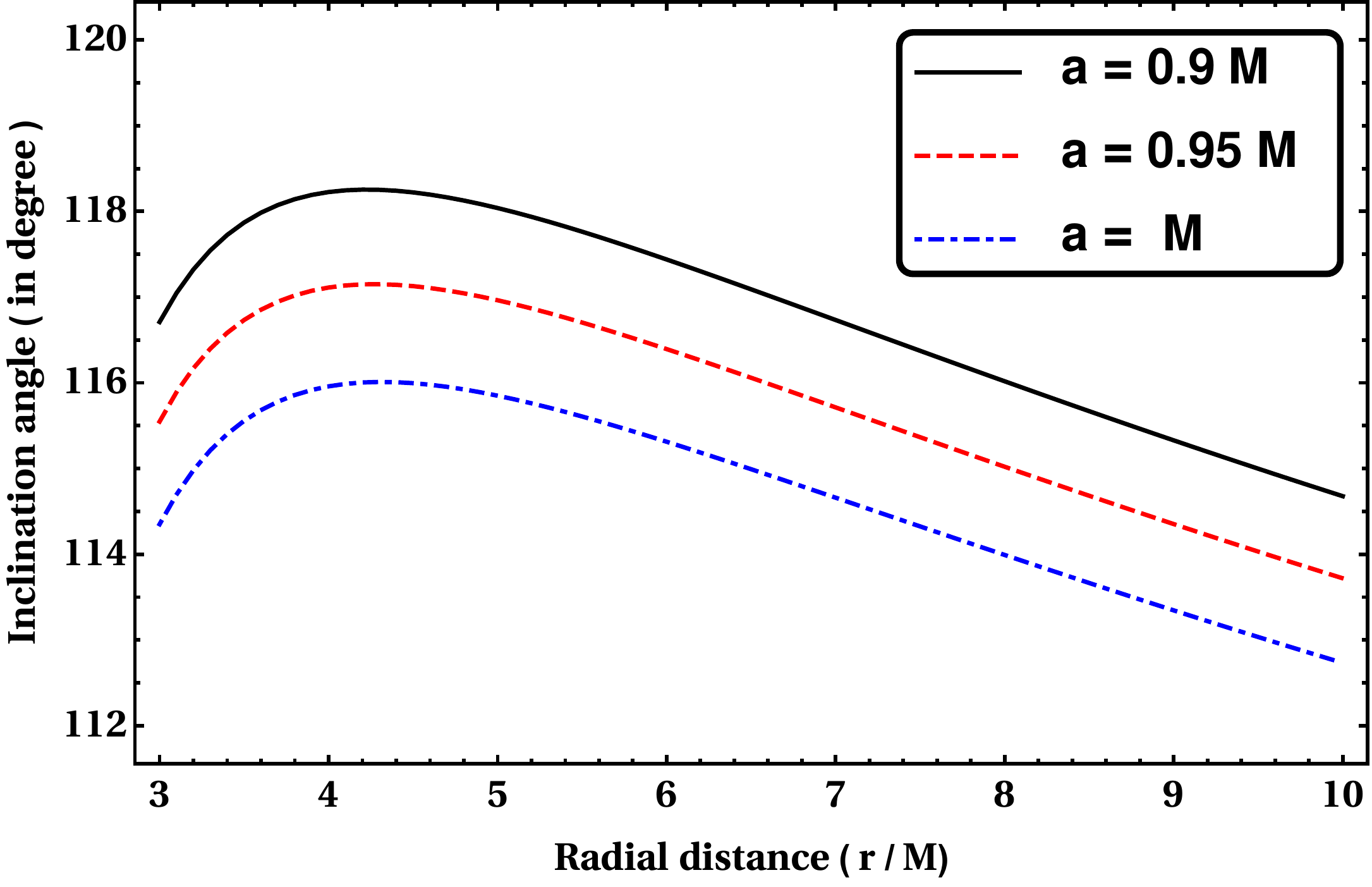}}
\hfill
\subfloat[Co-rotating circular orbits are given for $l=a$\label{Fig_7b}]{\includegraphics[height=5cm,width=.49\linewidth]{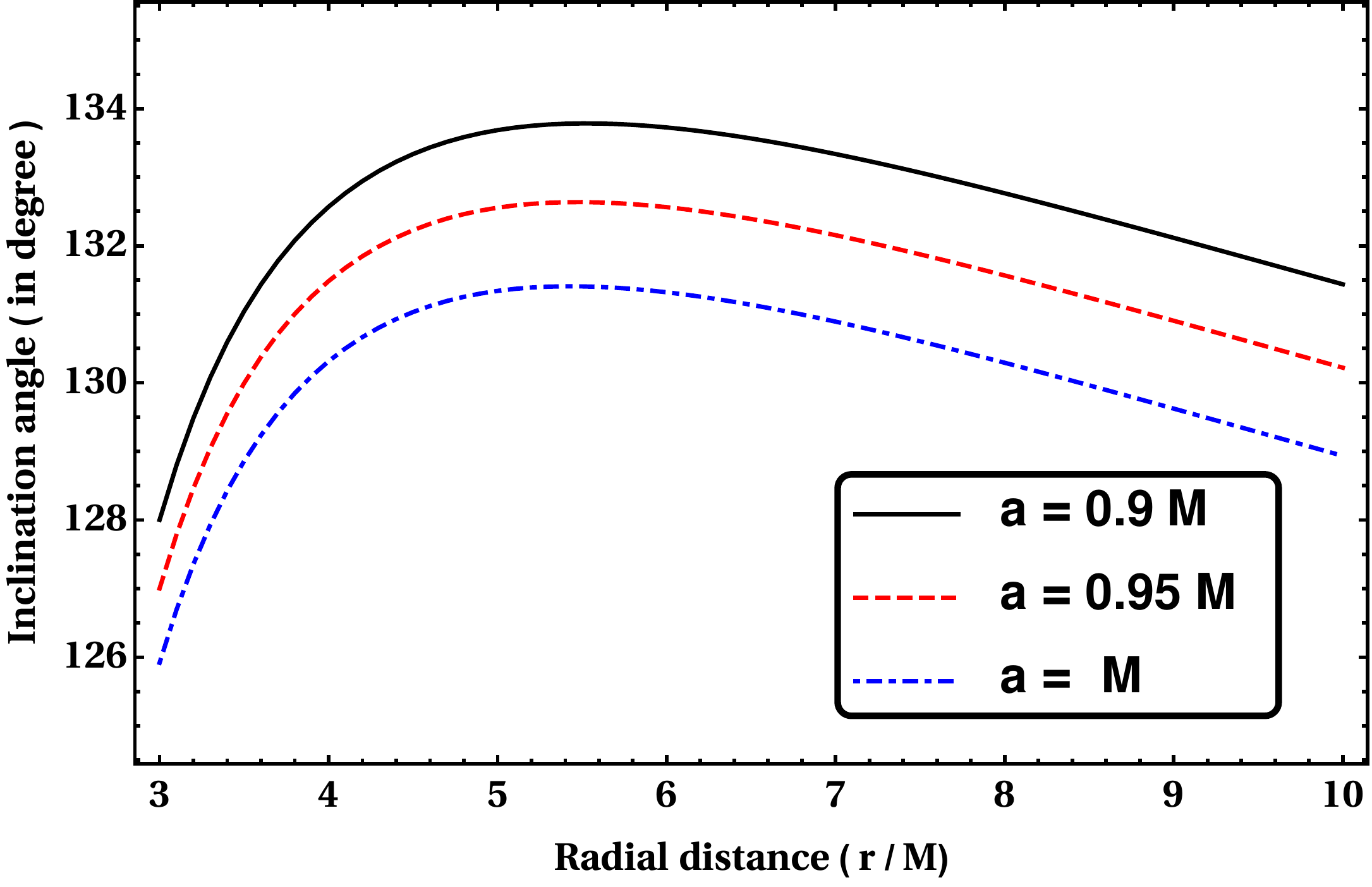}}
\caption{The angular dependence of the circular orbits are depicted in a \KNN~black hole satisfying $Q^2=l^2-a^2$.}
\label{Fig_7}
\end{figure}

\begin{figure}[htp]
\subfloat[For $l=a$, the counter-rotating circular orbits are shown for different rotational parameters of the black hole.  \label{Fig_8a}]{\includegraphics[height=5cm,width=.49\linewidth]{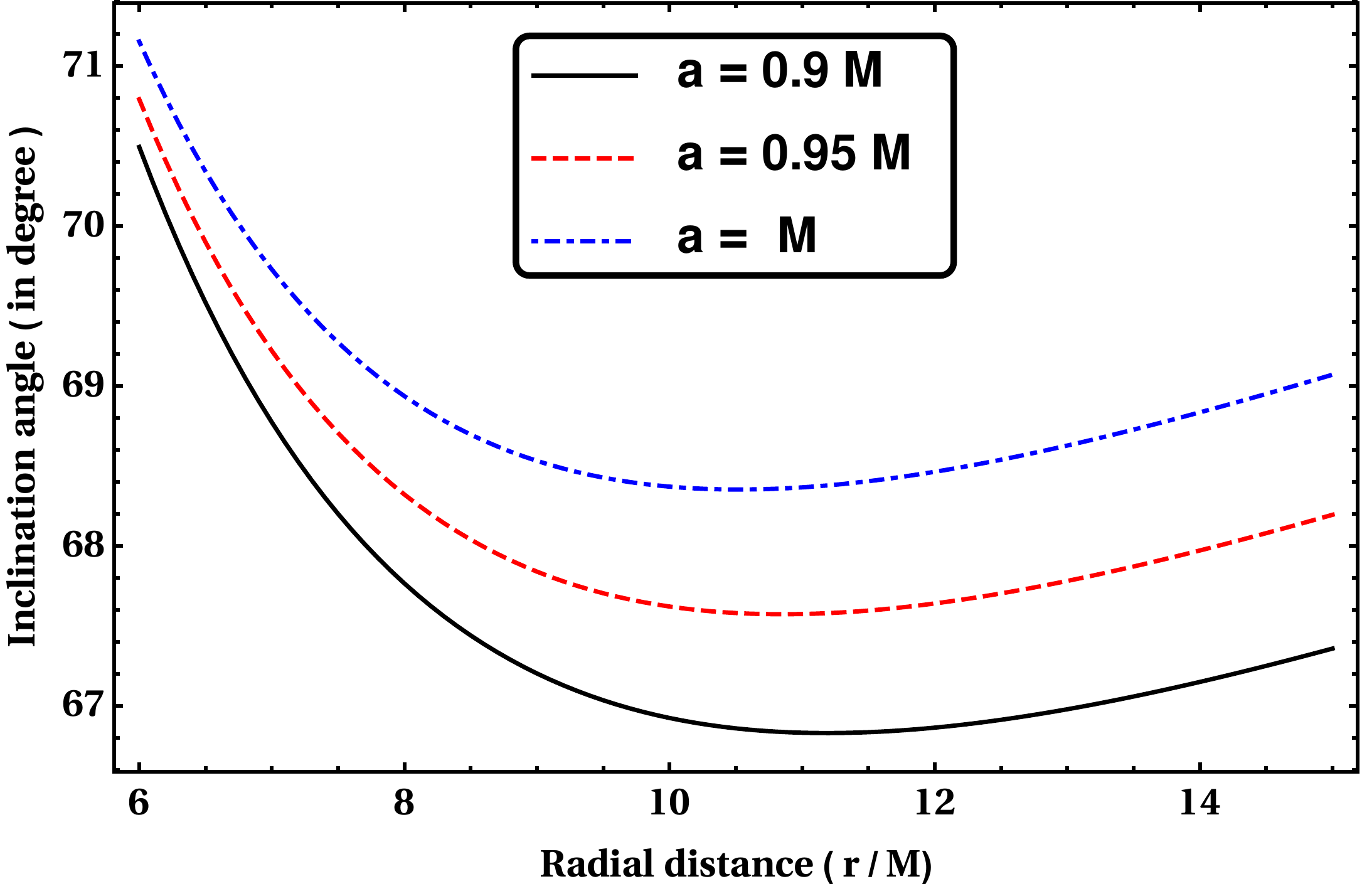}}
\hfill
\subfloat[Counter-rotating circular orbits are plotted for $l=2a$.\label{Fig_8b}]{\includegraphics[height=5cm,width=.49\linewidth]{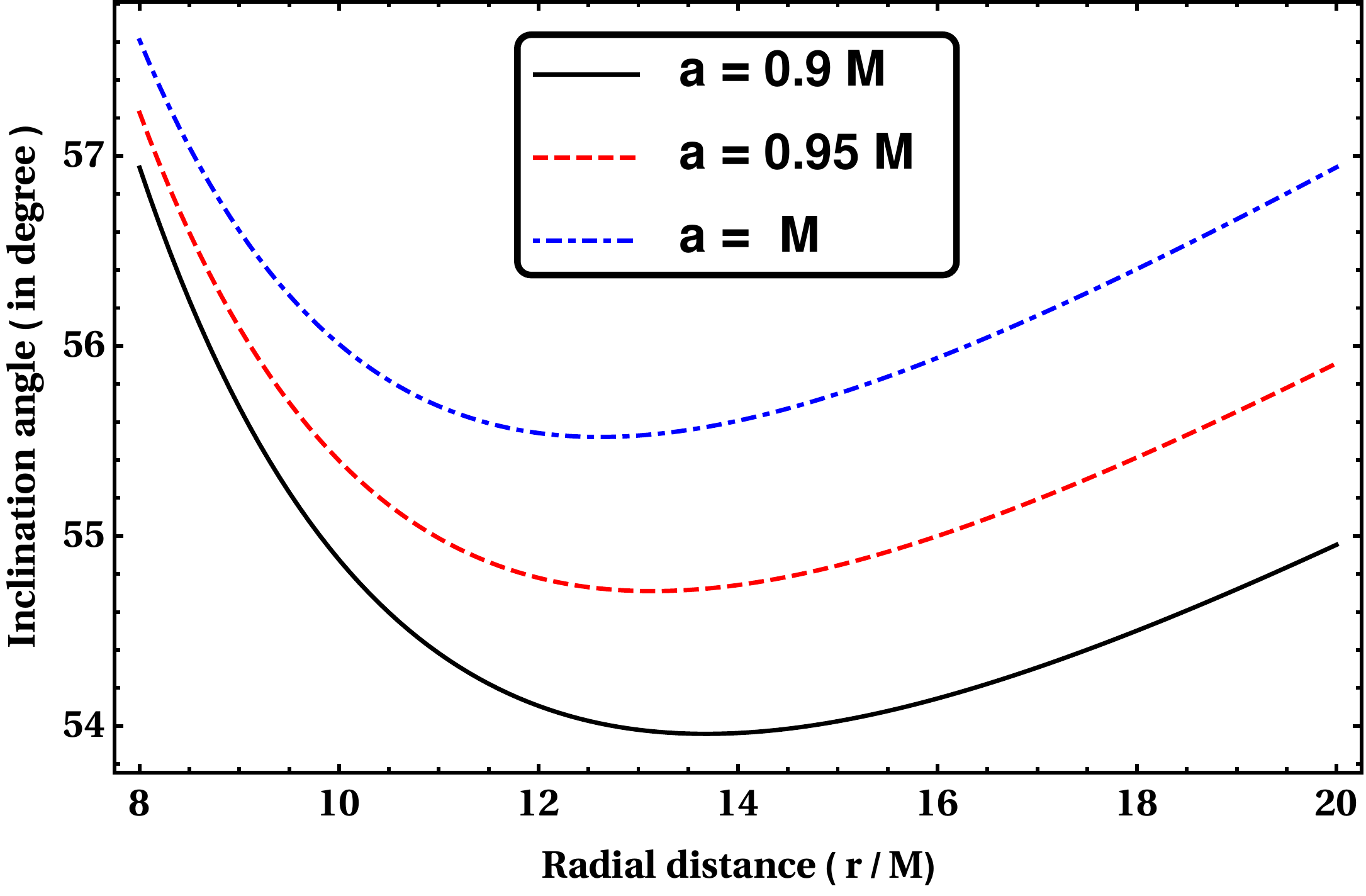}}
\caption{The counter-rotating circular orbits are shown for \KNN~blak hole with the horizon located at $r=2M$.}
\label{Fig_08}
\end{figure}

\noindent
With the given values for conserved energy and momentum, we can now address the angular constraints appear from $\dot{\theta}=\ddot{\theta}=0$. These two equations can be written further as
\begin{eqnarray}
{V}_{\rm eff}(\theta)&=& \left\{\lambda +(\tilde{L}_z-a\tilde{E})^2\right\} \sin^2\theta-(a\tilde{E} \sin^2\theta-2 \tilde{E} l \cos\theta-\tilde{L}_z)^2-\sin^2\theta(l+a \cos\theta)^2, \nonumber \\
{V}_{\rm eff}(\theta)&=& 2 \sin\theta \cos\theta \left\{\tilde{\lambda} +(\tilde{L}_z-a\tilde{E})^2\right\}-4(a\tilde{E} \sin^2\theta-2 \tilde{E} l \cos\theta-L_z)(a \tilde{E} \sin\theta \cos\theta+\tilde{E} l \sin\theta), \nonumber \\
&~~~~& -2 \sin\theta \cos\theta (l+a \cos\theta)^2+2 a \sin^3\theta (l+a \cos\theta).
\end{eqnarray}
By substituting the respective values for energy and momentum, we can write the above equations in the following form
\begin{equation}
V_{\rm eff}(\theta)=\mathcal{I}_{1}(a,l,r_{\rm c},\theta_c,\lambda_c), \quad \text{and} \quad \dfrac{dV_{\rm eff}(\theta)}{d\theta}=\mathcal{I}_{2}(a,l,r_{\rm c},\theta_c,\lambda_c),
\end{equation}
with $\mathcal{I}_1$ and $\mathcal{I}_2$ are two independent functions and there explicit forms can be both tedious and less illuminating, therefore excluded in the text. For a black hole given with specific NUT charge and rotation parameter, we can now numerically solve the above equations and locate the circular orbits confined in a particular plane. In \ref{Fig_7}, we have shown the co-rotating circular orbits for various momentum parameters of the black hole while the NUT charge is either $l=a$ (\ref{Fig_7a}) or $l=2a$ (\ref{Fig_7b}). Even the nature of figures remain identical, the deviation from the equatorial plane increases with the increase of the NUT charge. In case of the counter-rotating circular orbits as depicted in \ref{Fig_08}, the deviation from the equatorial plane is in exactly opposite direction from the co-rotating case and the inclination increases with an increase of the NUT charge. The role played by the rotational parameter of the black hole is also identical in both co-rotating and counter-rotating orbits. 

This finishes our discussion on trajectory of a massive as well as massless particle confined on a given plane. We will next take up the computation on the motion of a particle on an arbitrary plane.
\subsection{Non-Equatorial Plane}

In the previous section we have described the motion of both massive and massless particles in the equatorial plane of a \KNN\ black hole. However to understand some other subtle features associated with this spacetime it is important that we consider motion in non-equatorial plane as well. In this case, both for massive and massless particles the Carter constant will be non-zero and will play a significant role in determining various properties of the trajectory. 

Due to the complicated nature of the geodesic equations it will be worthwhile if we briefly recapitulate the non-equatorial trajectories of a massless particle for the Kerr-Newman black hole, which would be identical to Kerr geometry. In this case, the angular equation is given by:
\begin{equation}
\rho^4 \Big(P^{\theta}\Big)^2 + \cos^2\theta \left(\dfrac{L^2}{\sin^2\theta}-a^2 E^2\right) =\lambda.
\label{ang_Kerr}
\end{equation}
where $\lambda$ has the usual meaning of Carter constant and $P^{\theta}=d\theta/d\nu$ is the momentum in the $\theta$ direction with $\nu$ being the affine parameter along the null geodesic. From the above equation, one can easily read off the potential $V_{\rm ang}(\theta)$ associated with the angular motion as the second term on the left hand side of \ref{ang_Kerr}. Substituting a new variable $\mu=\cos\theta$, the angular potential can be written as \cite{o1995geometry},
\begin{equation}
V_{\rm ang}(\mu)=\dfrac{\mu^2}{1-\mu^2}\left\{L^2-a^2 E^2 (1-\mu^2)\right\} = \dfrac{\mu^2 a^2 E^2}{1-\mu^2}\left\{\mu^2-\left(1-\dfrac{L^2}{a^2 E^2}\right)\right\}.
\label{Vmu}
\end{equation}
Note that for $\mu=0$, or equivalently for $\theta_{0}=\pi/2$ the potential $V_{\rm ang}$ identically vanishes and the motion remains on the equatorial plane. The potential can be further sub-categorized by investigating the behaviour of the term, $L/aE$. For $L<aE$, the above equation vanishes at
\begin{equation}
\theta_1=\arccos \Big(-\sqrt{1-\frac{L^2}{a^2 E^2}}\Big);\qquad
\theta_2 =\arccos \Big(\sqrt{1-\frac{L^2}{a^2 E^2}}\Big)
\end{equation}
along with on the equatorial plane. It turns out that the values of $\theta$ for which the angular potential vanishes follow the order, $\theta_{2}<\theta_{0}<\theta_{1}$. On the other hand, for $L>aE$, it only vanishes on the equatorial plane. In what follows we will consider the case for which $L/aE<1$. In this case the above result suggests to rewrite the angular equation by introducing the two angles $\theta=\theta_{1}$ and $\theta =\theta _{2}$, such that,
\begin{equation}
\Big(\rho^{2} P^{\theta}\Big)^2  =\lambda-\dfrac{ a^2 E^2 (\cos\theta-\cos\theta_0)^2}{1-\cos^2\theta}\Big\{(\cos\theta-\cos\theta_1)(\cos\theta-\cos\theta_2)\Big\}
\end{equation}
It is obvious that for the momentum to have any real solution we must have the right hand side to be positive. With the following redefinitions: $\xi=(L/E)$, $\eta=(\lambda/E^2)$ and $\Theta=\rho^{4}\{P^{\theta}\}^2$, we arrive at the following expression for angular motion,
\begin{equation}
\Big(\dfrac{\Theta}{E^2}\Big)=\eta-\dfrac{ a^2 \cos^{2}\theta}{1-\cos^2\theta}\Big\{(\cos\theta-\cos\theta_1)(\cos\theta-\cos\theta_2)\Big\}.
\end{equation}
Let us now discuss three situations depending on the value of $\eta$, i.e., redefined Carter constant. We list below these three choices and associated physical implications.
\paragraph{(A) For $\eta=0$:}

The first case corresponds to a vanishing Carter constant. Since the Carter constant identically vanishes the momentum along the angular direction is negative of the angular potential. Thus for physical motion the angular potential should either be zero or negative.  In this case the following orbits are possible: 
\begin{figure}[htp]
\subfloat[We set $\xi=0.5M$.\label{Fig_1a}]{\includegraphics[height=5cm,width=.49\linewidth]{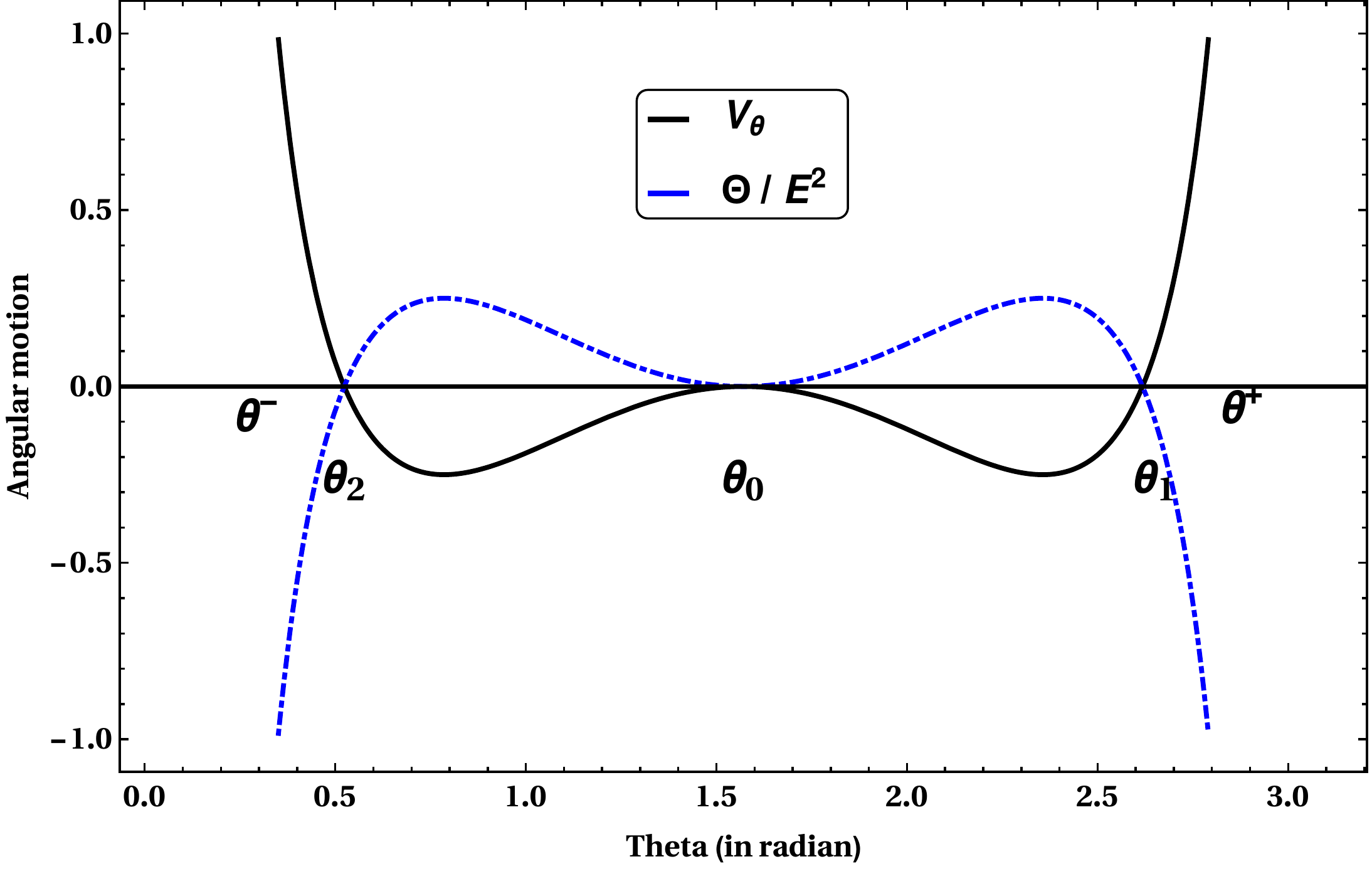}}
\hfill
\subfloat[We set $\xi=M$.\label{Fig_1b}]{\includegraphics[height=5cm,width=.49\linewidth]{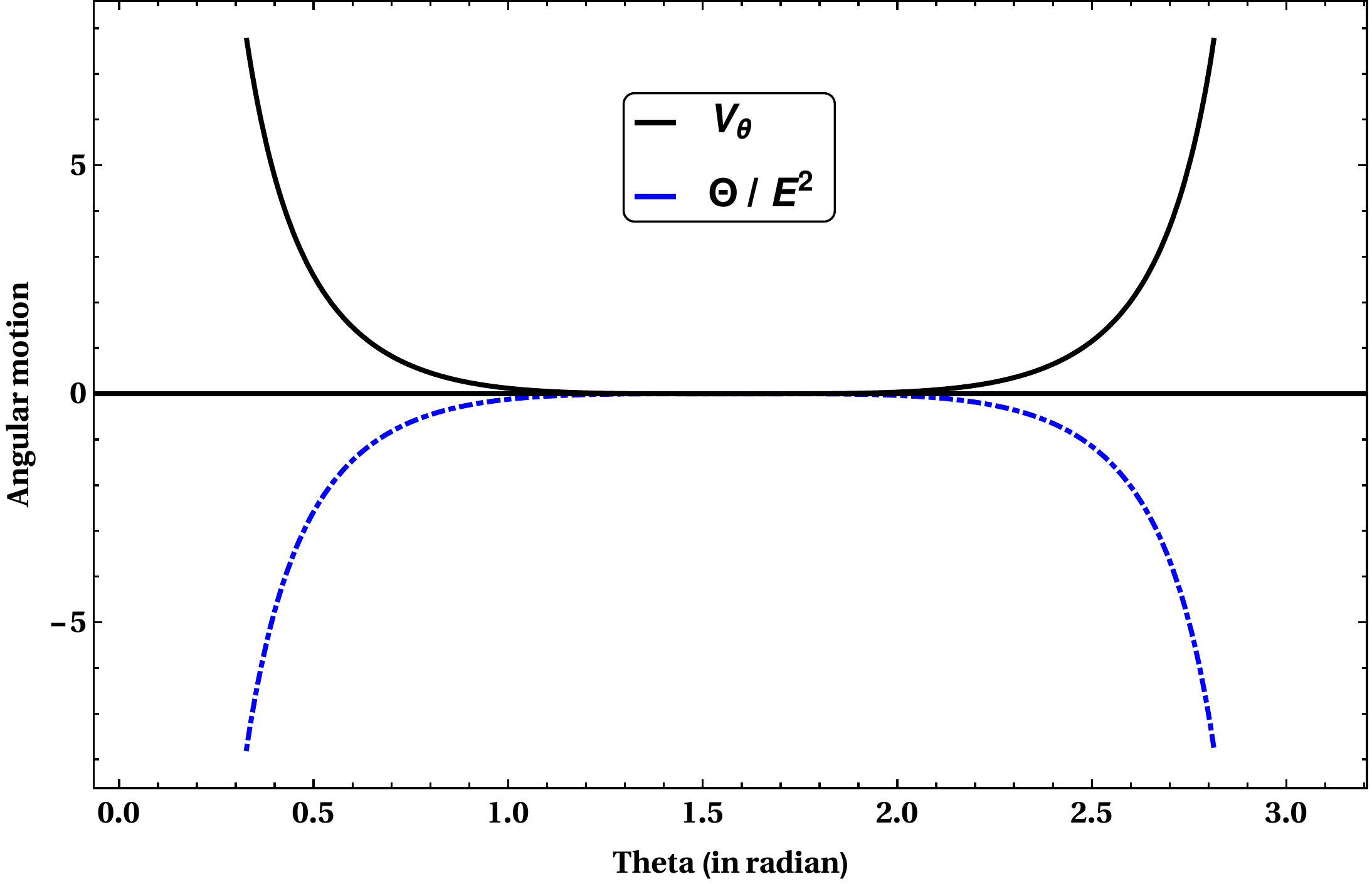}}
\caption{Angular potential and momentum for vanishing Carter constant (i.e., $\lambda=0$) has been presented for the following choice of the black hole rotation parameter: $a=M$.}
\label{Fig_1}
\end{figure}
\begin{itemize}

\item $P^{\theta}$ vanishes for $\theta=\theta_0$, $\theta_1$ and $\theta_2$ which are referred to as the turning points in the $\theta$ direction. In addition, the potential attains its local maximum value on the equatorial plane. Hence, a particle would remain on the equatorial plane unless acted upon by an external perturbation (see \ref{Fig_1a}), even though it is not a stable equilibrium point.

\item  Along with the confined motion on the equatorial plane, a massless particle can have off equatorial trajectories whenever $V(\theta)$ is negative. This is only possible if it follows, $\theta_2<\theta<\theta_0$ and $\theta_0<\theta<\theta_1$. The particle never reach the equator as it asymptotically approaching the equatorial plane. This is shown in \ref{Fig_1a}.

\item For $\xi=(L/E)=M$, one has $L/aE>1$ and hence in this case the potential has a minima and it vanishes \emph{only} on the equatorial plane. Since it never becomes negative, motion is only allowed on the equatorial plane with vanishing momentum along the angular direction (see \ref{Fig_1b}), i.e, $P^{\theta}=0$.

\end{itemize}

\paragraph{(B) For $\eta>0$:}

\begin{figure}[htp]
\subfloat[We set $\xi=0.5M$.\label{Fig_2a}]{\includegraphics[height=5cm,width=.49\linewidth]{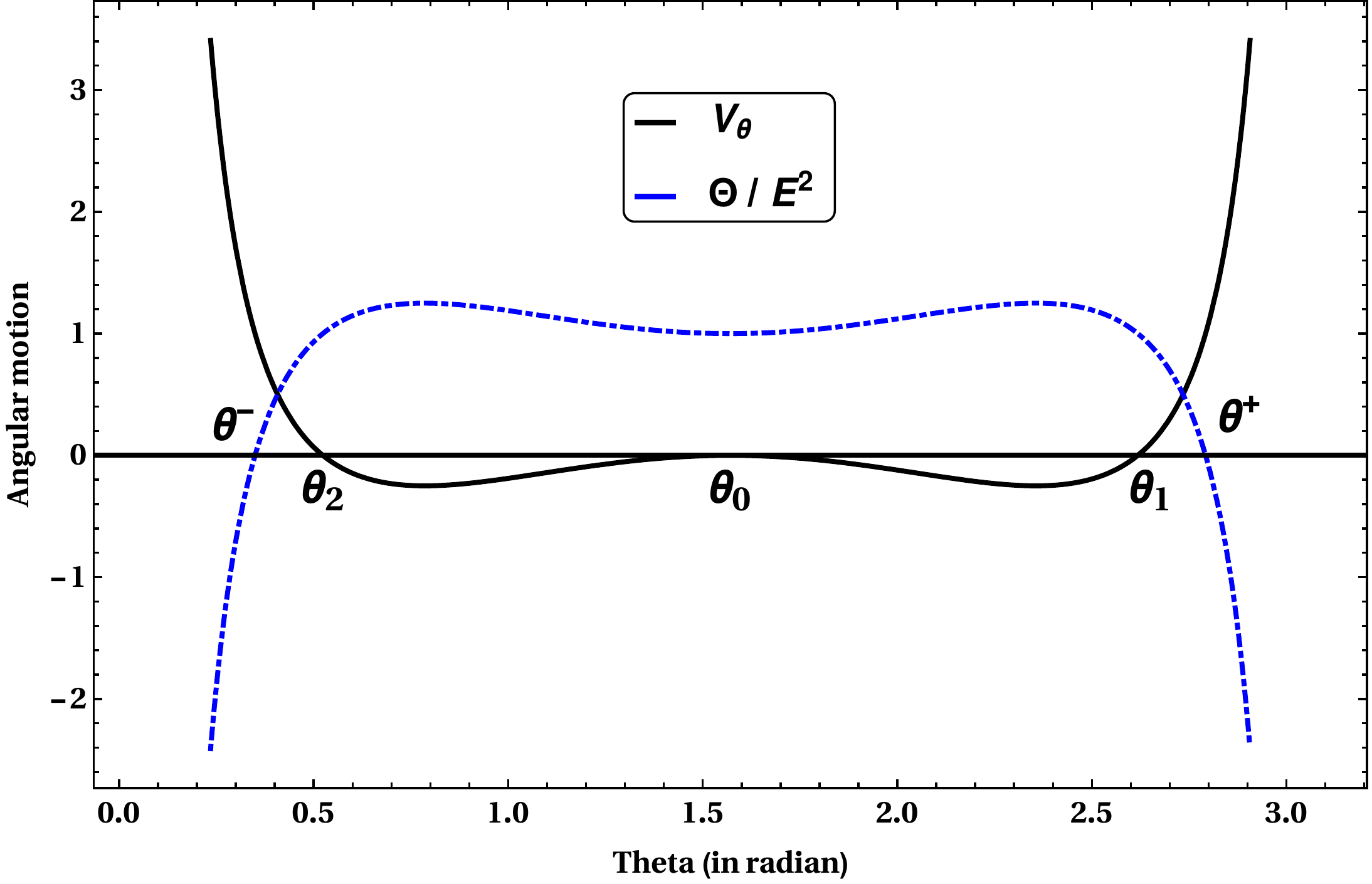}}
\hfill
\subfloat[We set $\xi=M$.\label{Fig_2b}]{\includegraphics[height=5cm,width=.49\linewidth]{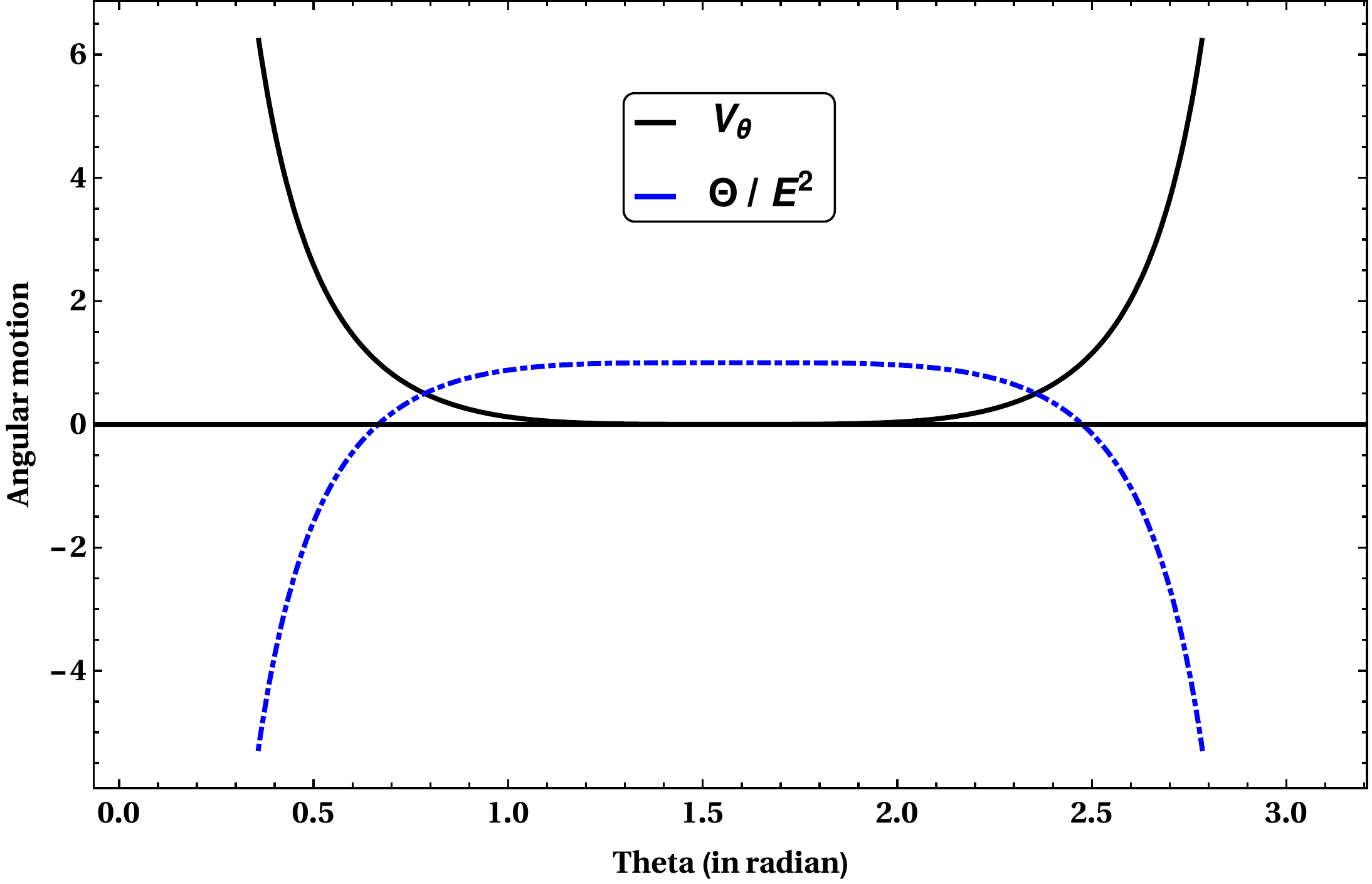}}
\caption{The angular potential and momentum associated with the angular motion of a massless particle in Kerr-Newman spacetime has been presented with non-zero Carter constant: $\lambda=M^2$ and rotation parameter: $a=M$.}
\label{Fig_2}
\end{figure}

In this case, the Carter constant is positive and hence there can be two possibilities as far as the angular potential is concerned. These include:
\begin{itemize}

\item For $(L/aE)<1$, the potential $V_{\rm ang}$ is negative within the region $\theta_2<\theta<\theta_1$. Since $\eta>0$ it follows that the momentum $P^{\theta}$ has two turning points located at $\theta ^{\pm}$, satisfying: $\theta _{-}<\theta _{2}$ and $\theta_{+}>\theta _{1}$ respectively. As evident from \ref{Fig_2a} the particle oscillates about the equatorial plane.

\item For $(L/aE)>1$, unlike the previous case, here the particle can travel beyond the equatorial plane upto the point where $\{P^{\theta}\}^{2}$ vanishes. Thus in this case the particle can travel away from the equatorial plane. This is depicted in \ref{Fig_2b}. 

\end{itemize}

\paragraph{(C) For $\eta<0$:} If the Carter constant becomes negative, then in order to ensure that $\{P^{\theta}\}^{2}$ is positive one must have negative $V_{\rm ang}$. In this case the following results are obtained:
\begin{figure}[htp]
\subfloat[We set $\xi=0.5M$.\label{Fig_3a}]{\includegraphics[height=5cm,width=.49\linewidth]{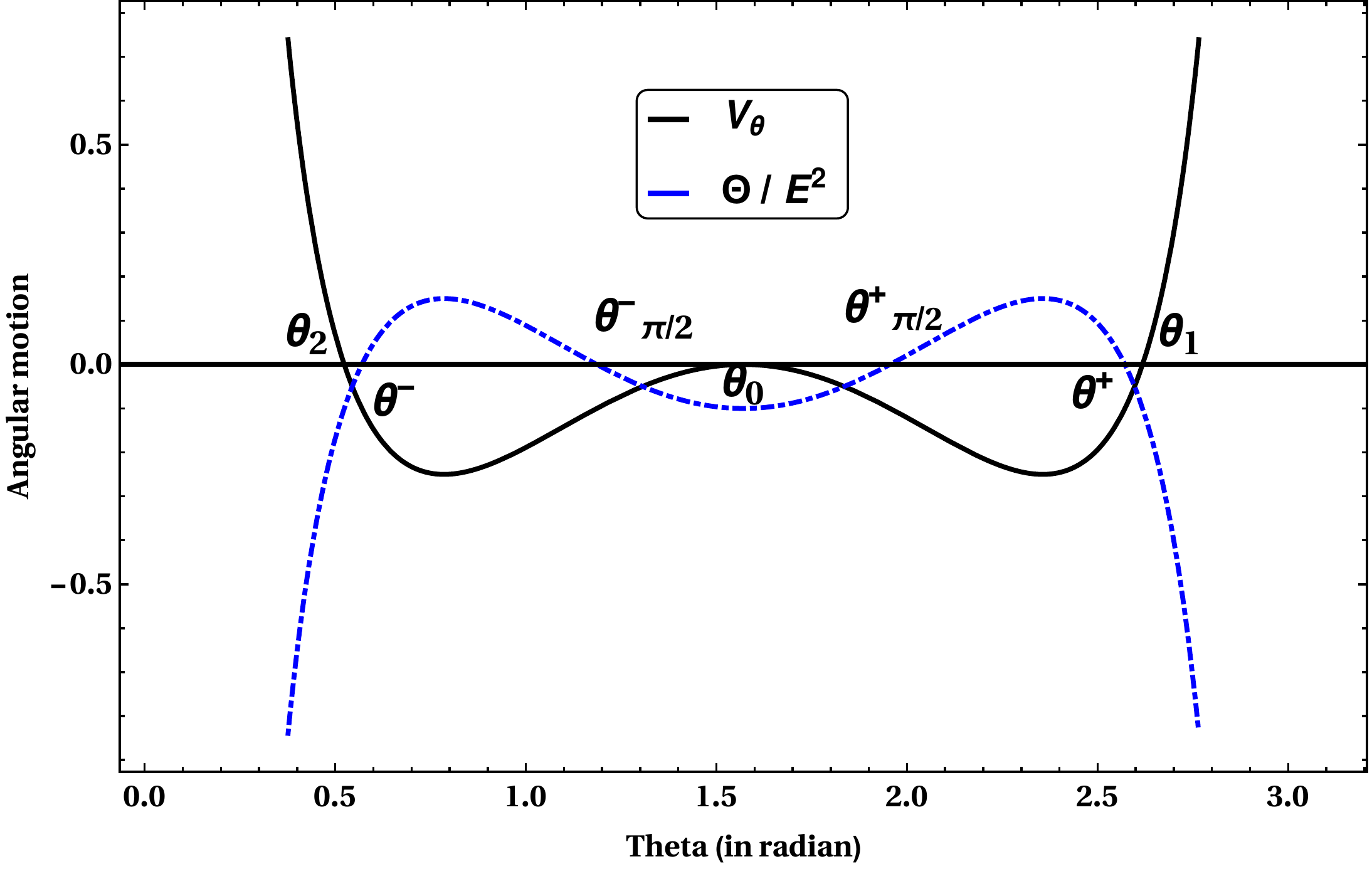}}
\hfill
\subfloat[We set $\xi=M$.\label{Fig_3b}]{\includegraphics[height=5cm,width=.49\linewidth]{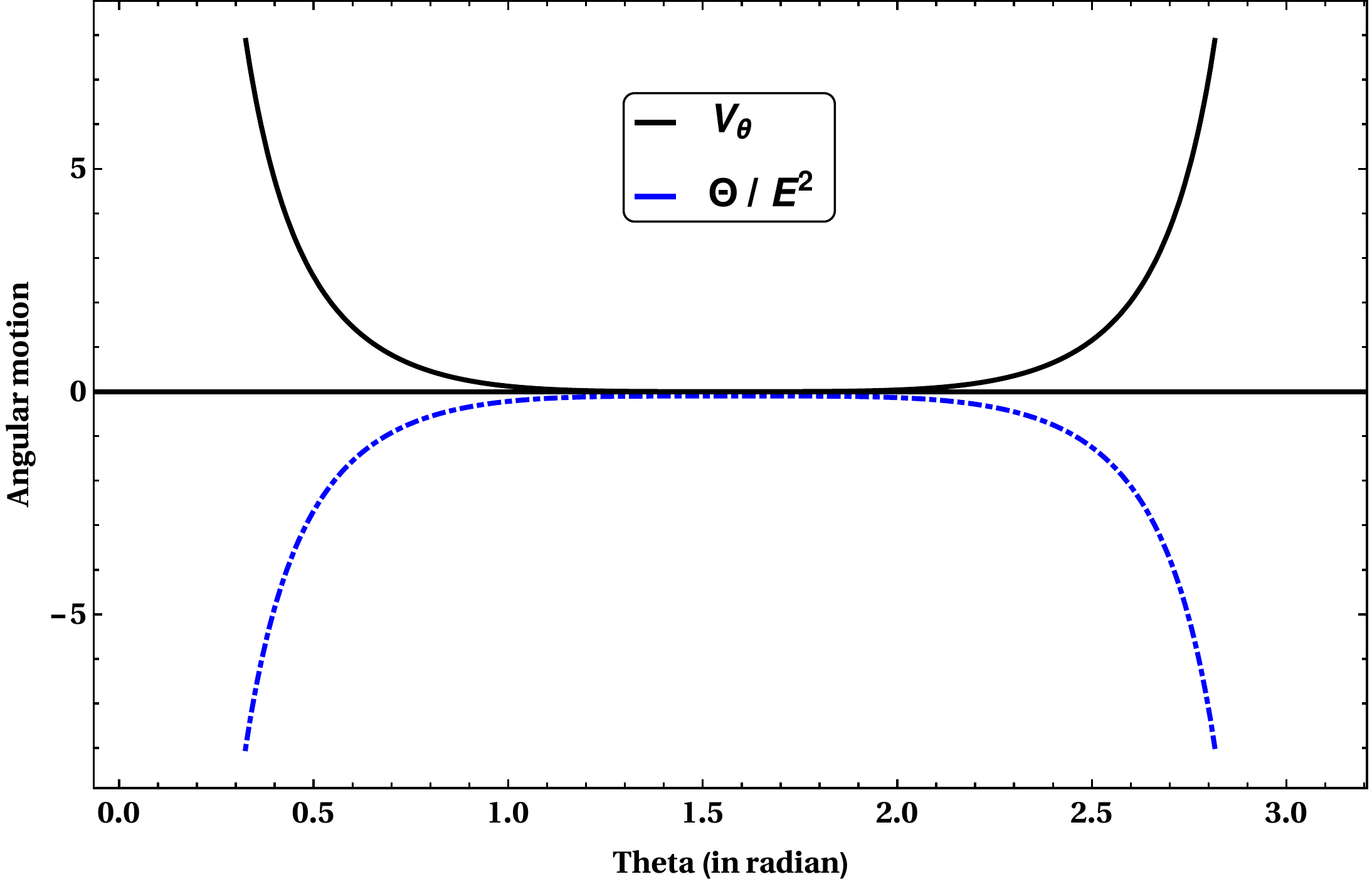}}
\caption{The momentum and potential associated with the angular motion of a massless particle in the Kerr-Newman spacetime is depicted with negative Carter constant, i.e., with $\lambda=-0.1M^2$ and rotation parameter: $a=M$.}
\label{Fig_3}
\end{figure}
\begin{itemize}

\item In this case with $(L/aE)<1$, not only $V_{\rm ang}$ has to be negative, one has to ensure that $|V_{\rm ang}|>|\lambda|$. Thus on the equatorial plane $V_{\rm ang}$ vanishes and hence there can be no physical motion on the equatorial plane. Thus in this case the particle has to travel off the equatorial plane. In particular, the particle has to obey either of these two conditions --- (a) $\theta_2<\theta^{-}<\theta <\theta_{\pi/2}^{-}<(\pi/2)$ or, (b) $(\pi/2)<\theta_{\pi/2}^{+}<\theta<\theta^{+}<\theta_1$, where $\theta_{\rm \pi/2}^{\pm}$ and $\theta^{\pm}$ are the turning points of the momentum $P^{\theta}$. A qualitative description can be found in \ref{Fig_3a}.

\item On the other hand, for $(L/aE)>1$ it follows that $V_{\rm ang}$ is always positive. Thus in this case there is absolutely no phase space available for the particle. Hence this corresponds to a unphysical situation, as presented in \ref{Fig_3b}.

\end{itemize}

\subsubsection{Massless Particles in \KNN~black holes}

Having described the trajectories of a massless particle in the context of Kerr-Newmann black hole as a warm up exercise, let us now present the orbits of a massless particle in the \KNN\ spacetime. Through this exercise we can easily read off the differences appearing due to the presence of NUT charge in the present context. In this case, by defining $R=\rho^{4}\{P^{r}\}^{2}$ and $\Theta=\rho^{4}\{P^{\theta}\}^{2}$, with $P^{r}=dr/d\nu$ and $P^{\theta}=d\theta/d\nu$, the following expressions for the radial and angular equations are obtained,
\begin{align}
\left(\frac{R}{E^{2}}\right)&=\left\{(r^2+a^2+l^2)-a\xi\right\}^2-\Delta (\xi-a)^2-\Delta \eta ,
\label{EOM_KNN_NonEqR}
\\
\left(\frac{\Theta}{E^{2}}\right)&=\eta-\Big\{(a\sin\theta-\xi\csc\theta-2 l\cot\theta)^2-(\xi-a)^2 \Big\}.
\label{EOM_KNN_NonEqT}
\end{align}
Here we have introduced the following notations, namely $\xi=L/E$ and $\eta=\lambda/E^2$. We will now consider the angular part of the geodesic equation before considering the radial part of the same. 
\paragraph{(A) angular motion:}

Similar to the previous case with Kerr-Newman black hole, given the angular equation one can write down the associated potential by substituting $\mu=\cos\theta$, resulting into
\begin{equation}
V_{\rm ang,gen}(\mu)=\dfrac{\mu}{1-\mu^2}\Big\{4 l \xi+4 l^2 \mu+\xi^2 \mu+4 a l (\mu^2-1)+a^2 \mu (\mu^2-1)\Big\}.
\label{Pot_ang_KNN}
\end{equation}
Thus the angular potential depends on quartic powers of $\mu$. Hence the angular coordinates where the potential vanishes correspond to a quartic equation for $\mu$. As evident from \ref{Pot_ang_KNN} the solution $\mu=0$ (or, $\theta=\pi/2$) is a trivial solution and hence even in this context the potential in the angular direction vanishes on the equatorial plane. Hence, one can rewrite the angular potential as,
\begin{equation}
V(\mu)=\dfrac{a^{2}\mu}{1-\mu^2}(\mu-\mu_1)(\mu-\mu_2)(\mu-\mu_3).
\end{equation}
Whether all the three solutions will be real or not depends on the choices of the parameter space spanned by $a$, $l$ and $\xi$ respectively. It turns out that one can have the two following  possibilities. 
\begin{itemize}
\item If the rotation parameter $a$, NUT charge $l$ and the specific angular momentum $\xi$ are such that the following condition is satisfied 
\begin{equation}
\mathcal{G}\equiv \left\{a^4-44 a^2 l^2+2 a^3 \xi -(4 l^2+\xi^2)^2 +a^2 (56 l^2 \xi-2 \xi^3) \right\}>0
\end{equation}
the solutions to \ref{Pot_ang_KNN} can be written as,
\begin{align}
\mu_{1}&=-\dfrac{4 l}{3 a}+\dfrac{(3 a^2+4 l^2-3\xi^2)^{1/2}}{3 a} \left\{2 \cos\alpha\right\}
\\
\mu_{2,3}&=-\dfrac{4l}{3a}-\dfrac{(3a^2+4 l^2-3 \xi^2)^{1/2}}{3a}(\cos\alpha \mp \sqrt{3}\sin\alpha).
\end{align}
Here $\alpha$ is an angle within the range $(0,2\pi)$ and can be explicitly written as
\begin{equation}
\alpha=\dfrac{1}{3}\arctan \left(\dfrac{3 \sqrt{3}B}{A}\right ),
\end{equation}
with $A$ and $B$ defined as, $A=36 a^5 l-54 a^4 l \xi+2 a^3 (4 l^3+9 l \xi^2)$ and $B=a^3(a-\xi)\sqrt{\mathcal{G}}$. For $l>0$, the order of the solutions for the angular variables are, $\theta_3>\pi>\theta_2 >\theta_0(=\pi/2) >\theta_1$. Here, $\theta_{i}=\cos^{-1}(\mu_i)$ with $i$ running from 0 to 3. Thus  depending on the value of the Carter constant, one can have different motion in the angular direction following \ref{EOM_KNN_NonEqT}. Note that for $l=0$, one arrives at $\alpha=\pi/6$ and hence one gets three solutions such as, $\mu=0,\pm \sqrt{1-(\xi/a)^2}$. This is consistent with the corresponding result for Kerr black hole. (Note that the angular motion is independent of the choice $Q^{2}=l^{2}-a^{2}$ and hence the results derived above will be applicable even in the general situation. This is why we have discussed the $l=0$ limit in the context of angular motion.)

\item On the other hand if we have, $\mathcal{G}<0$ there will be two solutions. One of them corresponds to the usual equatorial plane while the other one is at $\mu=\mu^{\prime}$ and given by 
\begin{equation}
\mu^{\prime}=-\dfrac{4l}{3a}+\dfrac{1}{3 a^2} \left\{(A+3\sqrt{3}B^{\prime})^{1/3}+\dfrac{a^2 (3 a^2+4 l^2-3 \xi^2)}{(A+3\sqrt{3}B^{\prime})^{1/3}}\right\},
\end{equation}
where, $A$ is already mentioned earlier and $B^{\prime}$ is given as, $B^{\prime}=a^3(a-\xi)\sqrt{-\mathcal{G}}$. Interestingly, for $l=0$, $\mu^{\prime}$ become zero with the constraint $\xi>a$. This matches with the results discussed earlier in the context of Kerr-Newman black hole.
\end{itemize}
Having determined the angular coordinates marking the vanishing of the angular potential $V_{\rm ang,gen}$, we can comfortably describe the trajectories related to positive, negative and vanishing Carter constant. This is what we discuss next. 
\begin{enumerate}

\item \textbf{For $\eta=0$:} In the case of vanishing Carter constant, as evident from \ref{EOM_KNN_NonEqT}, the potential $V_{\rm ang,gen}$ has to be negative. This results into the following behaviours: 
\begin{itemize}

\item  For $\mathcal{G}>0$, the angular potential can be negative only if $\theta_{1}<\theta<\theta_{0}$ or, $\theta_{2}<\theta<\pi$ (see \ref{Fig_4a}). On the other hand with $\mathcal{G}<0$, one has $\theta^{\prime}<\theta<\theta_0(\pi/2)$, where $\theta^{\prime}=\cos^{-1}(\mu^{\prime})$ (see \ref{Fig_4b}). 

\item Unlike the angular motion in the context of Kerr-Newman black hole, $V_{\rm ang,gen}$ has neither a maxima nor a minima on the equatorial plane located at $\theta=\theta_{0}=\pi/2$. The motion is depicted in \ref{Fig_4} for a particular set of parameters. 

\end{itemize}
\begin{figure}[htp]
\subfloat[The above figure is for: $l=M/4$, $\mathcal{G}>0$.\label{Fig_4a}]{\includegraphics[height=5cm,width=.49\linewidth]{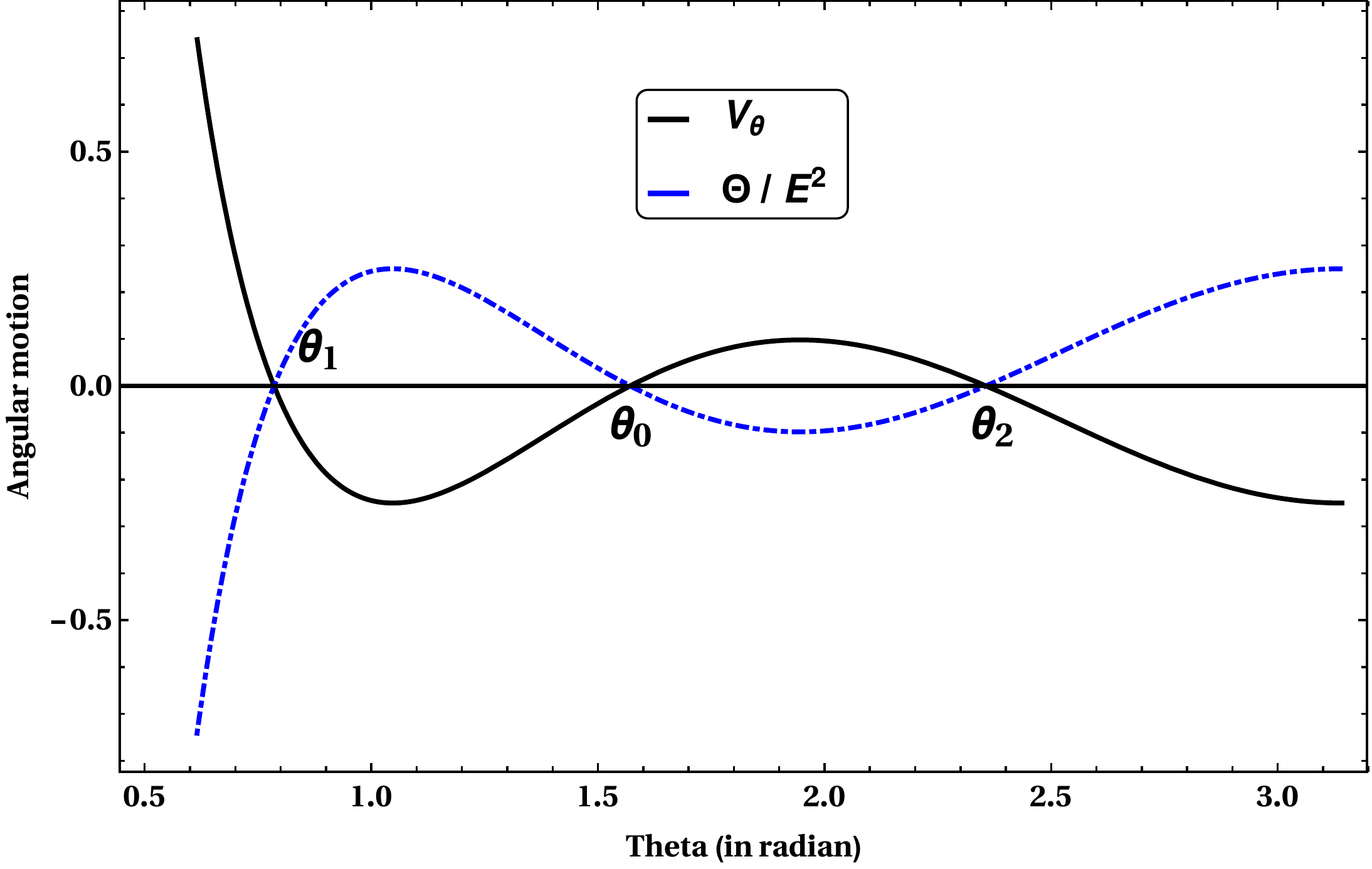}}
\hfill
\subfloat[In the above figure we have: $l=M$, $\mathcal{G}<0$.\label{Fig_4b}]{\includegraphics[height=5cm,width=.49\linewidth]{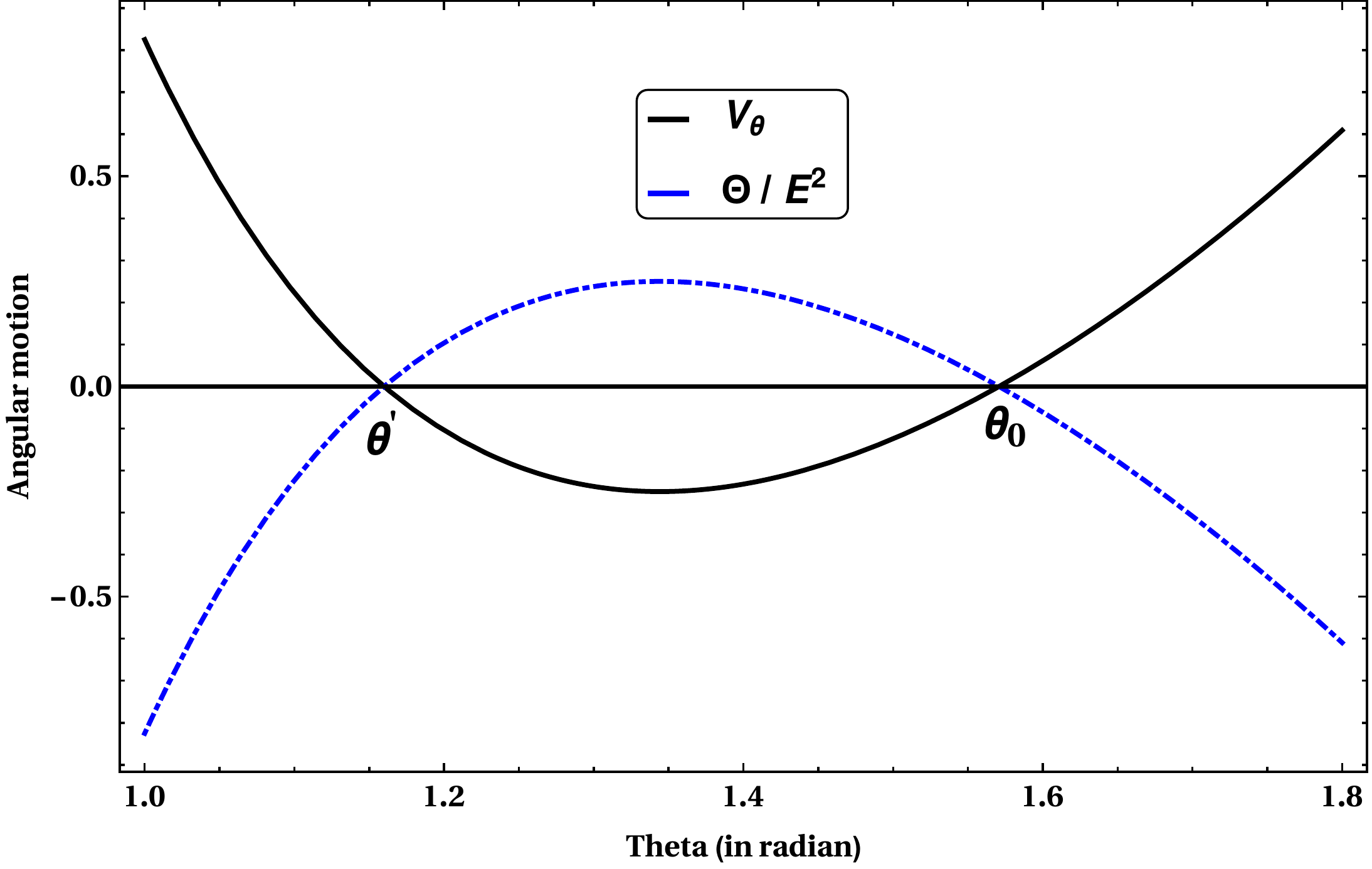}}
\caption{The angular potential and angular momentum for the \KNN\ black hole has been presented with vanishing Carter constant. Among other relevant quantities the impact parameter is taken as $L/E\equiv \xi=0.5M$ and finally the rotation parameter being $a=M$.}
\label{Fig_4}
\end{figure}

\item \textbf{For $\eta>0$:} In the case of positive Carter constant, the potential can take both positive and negative values. In the case of positive potential, motion along angular direction is possible only if numerical value of Carter constant is larger than the potential. 
\begin{itemize}

\item For positive values of $V(\theta)$ with $\mathcal{G}>0$, the particle has to be within the angular range: $\theta^{-}<\theta<\theta_{1}$ or, $\theta_{0}<\theta<\theta_{2}$ (see \ref{Fig_5a}). While for $\mathcal{G}<0$, orbits with exist, provided the angular coordinate satisfy: $\theta^{-}<\theta<\theta^{\prime}$ and $\theta_{0}<\theta<\theta^{+}$ (see \ref{Fig_5b}). Here $\theta^{\pm}$ are two turning points in the presence of positive Carter constant $\lambda$.

\item  $V(\theta)$ can also take negative values. In this case for $\mathcal{G}>0$, the only possibilities are: $\theta_{1}<\theta<\theta_{0}$ and $\theta_2<\theta<\pi$ (see \ref{Fig_5a}). Otherwise, with $\mathcal{G}<0$, one must have $\theta^{\prime}<\theta<\theta_{0}$ (see \ref{Fig_5b}).

\end{itemize}
\begin{figure}[htp]
\subfloat[In the above figure we have the following choices for the parameters: $l=M/4$, $\mathcal{G}>0$.\label{Fig_5a}]{\includegraphics[height=5cm,width=.49\linewidth]{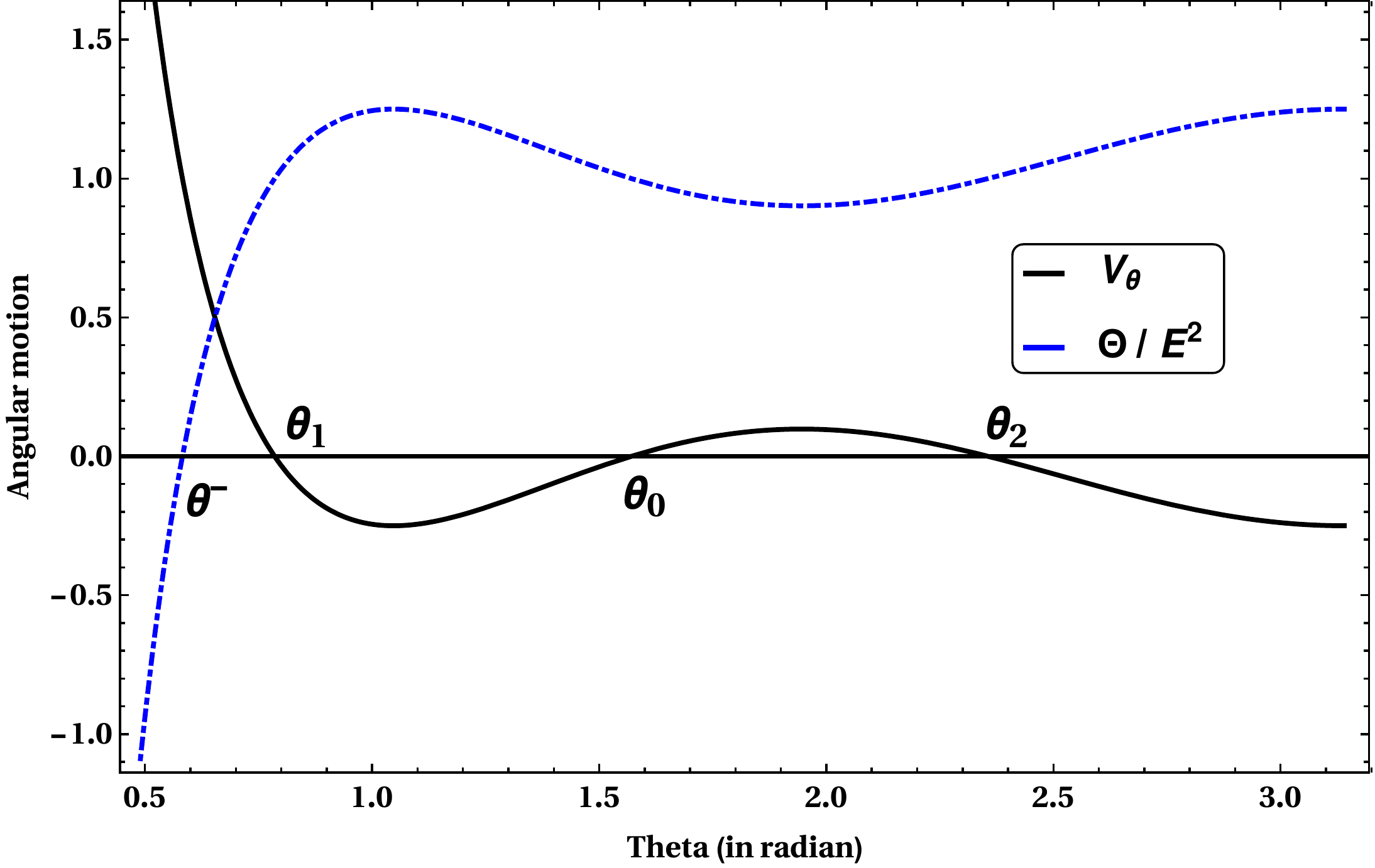}}
\hfill
\subfloat[The above figure is for $\mathcal{G}<0$ with $l=M$. \label{Fig_5b}]{\includegraphics[height=5cm,width=.49\linewidth]{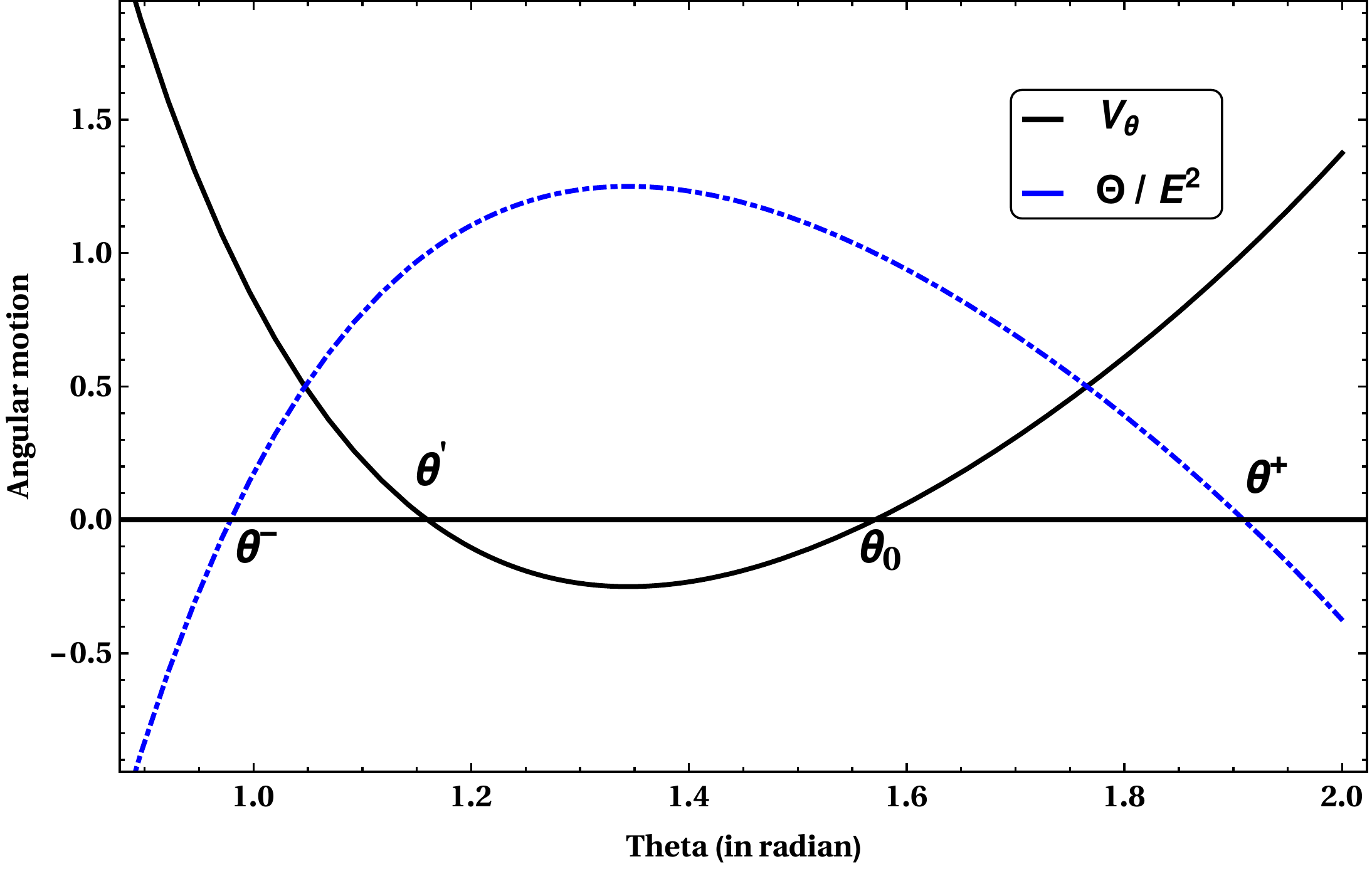}}
\caption{The angular motion of a particle in \KNN\ spacetime has been presented with a non-zero Carter constant $\lambda=M^2$, having the impact parameter $L/E=\xi=0.5M$ and the rotation parameter set to black hole mass.}
\label{Fig_5}
\end{figure}

\item \textbf{For $\eta<0$:} With negative Carter constant the potential can only be negative and also should have a magnitude larger than the Carter constant. This results into the following situation: 
\begin{itemize}

\item In this case, for $\mathcal{G}>0$, one has to ensure either $\theta_{1}<\theta^{-}<\theta<\theta_{\rm \pi/2}^{-}<\theta_0$ or $\theta_2 <\theta_{\rm \pi/2}^{+}<\theta<\pi$. Here $\theta^{\pm}$ and $\theta_{\rm \pi/2}^{\pm}$ are turning points of the momentum along the angular direction (see \ref{Fig_6a}). 

\item For $\mathcal{G}<0$, we need to have $\theta^{\prime}<\theta^{-}<\theta<\theta^{+} <\theta_0(\pi/2)$. The corresponding situation is depicted in \ref{Fig_6b}.
 
\end{itemize}
\begin{figure}[htp]
\subfloat[In the above we have set $l=M/4$, $\mathcal{G}>0$.\label{Fig_6a}]{\includegraphics[height=5cm,width=.49\linewidth]{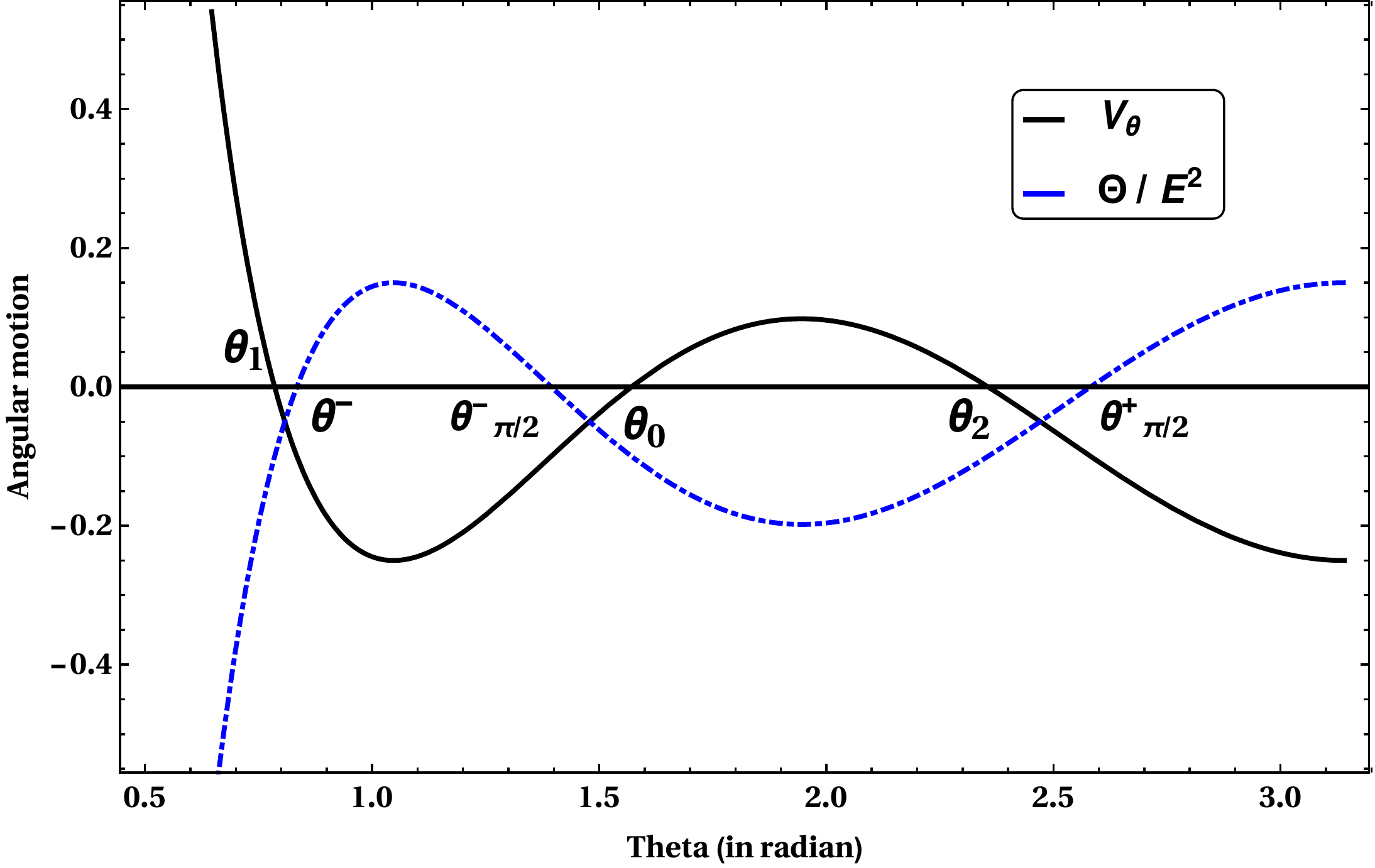}}
\hfill
\subfloat[The above figure is for the following parameter choice: $l=M$ along with $\mathcal{G}<0$.\label{Fig_6b}]{\includegraphics[height=5cm,width=.49\linewidth]{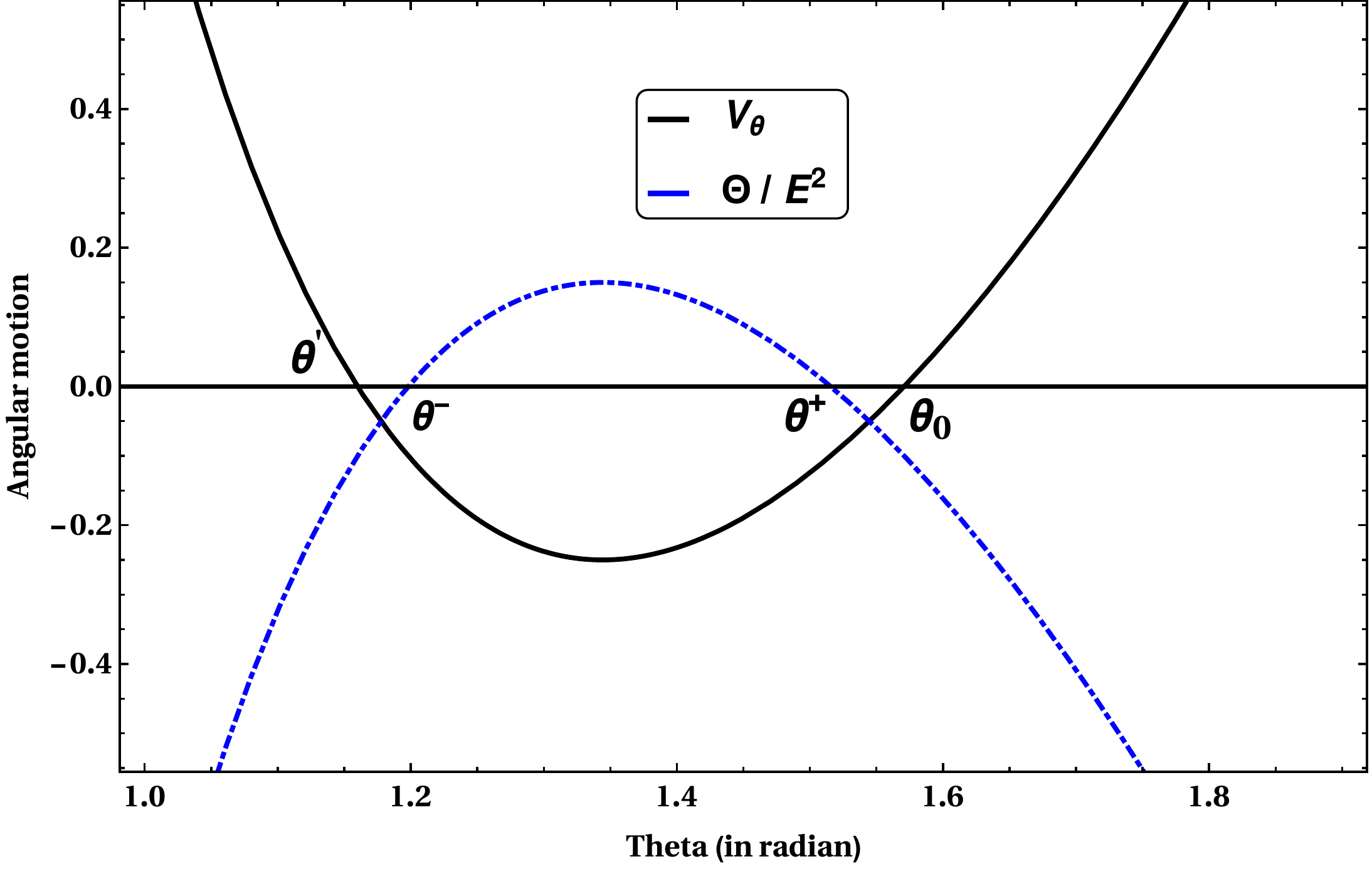}}
\caption{We have presented the potential and momentum responsible for angular motion of a massless particle with $L/E\equiv \xi=0.5M$, $a=M$ and $\lambda=-M^2/10$.}
\label{Fig_6}
\end{figure}

\end{enumerate}

\paragraph{(B) Radial Equation:}

The radial equation has already been presented in \ref{EOM_KNN_NonEqR} involving $\Delta$. In the present context of \KNN\ spacetime, the corresponding metric elements read, $\Delta=r^2-2Mr$. For the existence of a circular photon orbit, the necessary condition being $R=dR/dr=0$. Solving for $\eta=(\lambda/E^{2})$ and $\xi=(L/E)$, we arrive at two distinct solutions for the pair as, 
\begin{align}
\xi_c^{(1)}=\dfrac{r^2+l^2+a^2}{a};\qquad 
\eta_c^{(1)} =-\dfrac{(r^2+l^2)^2}{a^2}.
\label{Non_Eq_Radial_1}
\end{align} 
and,
\begin{align}
\xi_{c}^{(2)}&=\dfrac{a^2 (M-r)+l^2 (M-r)+r^2 (r-3M)}{a(M-r)}~,
\nonumber
\\
\eta_{c}^{(2)}&=-\dfrac{1}{a^2(r-M)^2}\Big\{l^4 (r-M)^2-2 l^2 r^2(r^2-4 M r+3 M^2)
\nonumber
\\
&\hspace{2cm}+r^3 \Big(r(r-3M)^2-4a^2(r-2M)\Big)\Big\}~.
\label{Non_Eq_Radial_2}
\end{align}
Considering the two solutions presented above for the parameters $\xi_c$ and $\eta_c$, one can explicitly demonstrate that in the first case (presented in \ref{Non_Eq_Radial_1}) $\eta _{c}+(\xi_{c}-a)^2$ become null and hence the angular equation will lead to $\Theta<0$. Thus for the first choice $(\xi_{c}^{(1)},\eta _{c}^{(1)})$ no angular motion is possible and hence we shall take \ref{Non_Eq_Radial_2} as the parameters associated with photon circular orbits. 
\subsubsection{Massive Particles in \KNN~black holes}

\paragraph{(A) Angular Equations:} For massive particles the geodesic equation associated with angular motion takes the following form, 
\begin{equation}
\left(\frac{\Theta_{\rm m}}{E^{2}}\right)=\eta-\Big\{(a\sin\theta-\xi\csc\theta-2 l\cot\theta)^2+\dfrac{m^2}{E^2}(l+a \cos\theta)^2-(\xi-a)^2 \Big\}
\end{equation}
Here $\Theta _{\rm m}=\rho ^{4}m^{2}(d\theta/d\tau)^{2}$ and the various quantities used in the above equation has the following definitions: $\eta \equiv \lambda/E^{2}$, $\xi \equiv L/E$, where $\lambda$ is the Carter constant. Due to complicated nature of the angular equations let us consider a simple situation with $E=m$. This corresponds to a marginally bound orbit and it is possible to analytically compute the nature of trajectories in the angular direction. In this context the above equation for angular motion reads,
\begin{equation}
\left(\frac{\Theta_{\rm m}}{E^{2}}\right)=\eta-\Big\{(a\sin\theta-\xi\csc\theta-2 l\cot\theta)^2+(l+a \cos\theta)^2-(\xi-a)^2 \Big\}
\label{massive_theta}
\end{equation}
Similar to the case for massless particles, there can be two possibilities, which will be discussed below. 
\begin{itemize}

\item In the first case, where we have all the solutions for $\theta$ originating from setting $\Theta _{\rm m}=0$ in \ref{massive_theta}, it is necessary that the parameters associated with the black hole satisfies the following identity,
\begin{align}
\mathcal{F}=\Big\{16 a^4 l^2-96 a^3 l^2  \xi 
&-3 \left(l^2-\xi^2\right)\left(3 l^2+\xi^2\right)^2
\nonumber
\\
&+a^2 \left(-72 l^4+180 l^2 \xi^2+\xi^4\right)+4 a \left(18 l^4 \xi-29 l^2 \xi^3-\xi^5\right)\Big\}>0
\end{align}
Unlike the case for massless particles, in the present context the angular potential does \emph{not} vanish at $\theta=\pi/2$, as evident from \ref{massive_theta}. However the angular potential being a cubic expression in terms of $\cos \theta \equiv \mu$, there will be at most three solutions for which the potential vanishes. The first solution takes the following form,
\begin{equation}
\mu_{1}=-\dfrac{3 l^2+\xi^2}{6 a l}+\dfrac{\chi}{3 a l}\cos\beta.
\label{eq:Equt_l}
\end{equation}
where $\chi^2\equiv 12 a^2 l^2-24 a l^2 \xi+(3 l^2+\xi^2)^2$ and $\beta$ is an angle taking values within the range $(0,2\pi)$ whose explicit expression can be given by,
\begin{equation}
\beta=\dfrac{1}{3}\arctan\Big(\dfrac{6\sqrt{3}\mathcal{D}}{\mathcal{C}}\Big),
\end{equation}
Here $\mathcal{D}$ and $\mathcal{C}$ are mathematical quantities having the following expressions,
\begin{equation}
\mathcal{C}=-27 l^6-27 l^4 \xi^2-9 l^2 \xi^4-\xi^6+36 a l^2 \xi (3 l^2+\xi^2)-18 a^2 l^2 (6 l^2+\xi^2),~\text{and}~\mathcal{D}=a l^2 \sqrt{\mathcal{F}}.
\end{equation}
The equation $\Theta _{\rm m}=0$ is a cubic equation for $\mu=\cos\theta$ and there should be three independent solutions at most. One of the solution is given above by $\mu=\mu_{1}$, while the other two solutions take the following form,
\begin{equation}
\mu_{2,3}=-\dfrac{3 l^2+\xi^2}{6 a l}-\dfrac{\chi}{6 a l}\left(\cos\beta \pm \sqrt{3}\sin\beta\right).
\end{equation}
The parameters introduced above have their usual meanings. 

\item The other possibility corresponds to $\mathcal{F}<0$. In this case two of the three solutions does not exist as they become complex. The only angular coordinate $\theta'$, where the potential associated with angular motion of a geodesic vanishes corresponds to, 
\begin{equation}
\cos \theta'\equiv \mu^{\prime}=-\dfrac{3 l^2+\xi^2}{6 a l}-\dfrac{1}{6 a l}\Big \{\Big(\mathcal{C}+6 \sqrt{3} \mathcal{D}^{\prime}\Big)^{1/3}+\dfrac{12 a^2 l^2-24 a l^2 \xi+(3 l^2+\xi^2)^2}{(\mathcal{C}+6 \sqrt{3}\mathcal{D}^{\prime})^{1/3}}\Big \}
\end{equation}
The quantity $\mathcal{C}$ appearing in the above equation has already been defined while, $\mathcal{D}^{\prime}=a l^2 (-\mathcal{F})^{1/2}$. Thus in this case there is only a single angular coordinate where the potential vanishes. This is certainly different from the corresponding situation in Kerr-Newman spacetime.

\end{itemize}
In passing, we would like to point out that for $l=0$ and $E=m$, the timelike geodesic has a vanishing angular potential only in the equatorial plane of a Kerr-Newman black hole. However, in our context with $l=0$, we have $\beta=(\pi/3)$ and substituting it back into \ref{eq:Equt_l}, we easily get $\mu_1=0$. On the other hand, solutions like $\mu_2$, $\mu_3$ and $\mu^{\prime}$ are only possible for a nonzero NUT charge and are discerning features of the \KNN\ spacetime.  

\paragraph{(B) Radial Equations:} In the case of radial orbits for massive particle, the corresponding equation is given by \ref{EOM_Radial}, which can also be written in the following form, 
\begin{equation}
\left(\frac{R}{E^{2}}\right)
=\left\{(r^{2}+a^{2}+l^{2})-a\xi\right\}^{2}-\left(\eta+\frac{m^{2}}{E^{2}}r^{2}\right)\Delta -\Delta \left(\xi-a\right)^{2} 
\end{equation}
In the above expression we have $R=m^{2}{\rho ^{4}}(dr/d\tau)^{2}$ and rest of the quantities have their usual meaning. Alike the case for a massless particle, the motion of a massive particle also involves circular orbits, which can be obtained by setting both $R$ and its radial derivative to zero. This leads to the following choices for the impact parameter $\xi\equiv L/E=\xi_{\rm c}$ and the effective Carter constant $\eta \equiv \lambda/E^{2}=\eta _{c}$, such that for a marginally bound orbit with $E=m$, we arrive at
\begin{align}
\xi_{\rm c}&=\dfrac{1}{a(M-r)}\left\{M(r^2-a^2)+a^2 r+l^2 (r-M) \pm r \Delta \sqrt{\dfrac{M}{r}} \right\}, 
\nonumber 
\\
\eta_{\rm c}a^2(r-M)^{2}&=\left\{M r^2 a^2 (r-3 M)-\Big[M r^3 (r^2-3 M r+4 M^2)+2 M r^2 l^2 (r-M)+l^4 (r-M)^2\Big]\right\}.
\nonumber 
\\
&\mp2 \Delta \sqrt{M r}\Big[Mr^2+l^2 (r-M)\Big] \mp 2 a^2 r \Delta \sqrt{M r}. 
\end{align}
In the above expressions for $\xi_{\rm c}$ and $\eta _{\rm c}$, the radial coordinate appearing on the right hand side corresponds to the location of circular orbits. For vanishing Carter constant, one can solve the equation involving $\eta _{\rm c}$ to write down the energy in terms of radius of the circular orbit. This result when used in the expression for impact parameter $\xi_{\rm c}$ will also help in expressing the angular momentum in terms of the radius of circular orbits. Thus circular orbit on the non-equatorial plane can exist for a given radius $r$ and given energy $E_{\rm c}$, provided the effective Carter constant $\eta _{\rm c}$ and impact parameter $\xi_{\rm c}$ are given by the above expressions.
\section{Energy extraction from Kerr-Newman-NUT black hole}
\label{Energy_KNN}

The origin of high energy particles in the universe is a long standing problem. Even though, there have been several explorations to model such phenomena \cite{LetessierSelvon:2011dy,Hillas:1985is}, it can be theoretically intriguing if it has its roots back to some exotic objects, such as black hole or neutron star. Historically, many high energetic events in the universe has their connections one way or another into black hole or stars, such as formation of jets from rotating objects as a result of gamma ray burst \cite{piran2001energy} or active galactic nuclei \cite{sauty2002jet}. More recently, it has been proposed that black holes could also be used as a system to accelerate particles, giving rise to arbitrary large energy Debris \cite{Berti:2014lva,Schnittman:2014zsa,Harada:2014vka}, which in principle dictates a modified version of the Penrose process. This idea was originally suggested by Penrose and Floyed in the late seventies \cite{penrose1971extraction}, concerning energy extraction from a black hole in the presence of a ergoregion. Since then, many investigations have been carried out in many aspects to examine the implications of Penrose process in various astrophysical domains \cite{Patil:2015fua,Leiderschneider:2015kwa,Bejger:2012yb,Armaza:2015eha,Mukherjee:2018kju,Maeda:2018hfi,Liu:2018myg,Dadhich:2018gmh,wagh1985revival,Blandford:1977ds,Mukherjee:2018cbu}. 

In the present context, we would reconsider the possibilities of energy extraction from a \KNN~black hole constrained with $\Delta=r^2-2Mr$ and $Q^2=l^2-a^2$. We start with the original Penrose process and study the bounds from Wald inequality. Afterward, we investigate the implication of \CPP~followed by a survey of recent Ba\~{n}ados-Silk-West effect regarding the divergence of collisional energy in the center of mass frame. Finally, we will address the superradiance phenomenon in the \KNN~black hole and discuss the advantages over other spacetimes such as Kerr or Kerr-Newman.

\subsection{The original Penrose process}
\label{Penrose_Process}

In the original Penrose process, the idea is to send a particle that breaks up into two in the ergoregion, one of which crosses the horizon with a negative energy while the other comes out with energy more than the initial energies. By this means, rotational energy of a black hole could be extracted. For this purpose, the energy of the initial particle plays a crucial role. Thus we would like to write the energy of the particle in terms of angular momentum. This can be achieved by setting either $P_{r}=0$ or by substituting all the momentum components in the on-shell condition, i.e., $P_{\mu}P^{\mu}=-m^{2}$. They both should give identical result. Thus substituting $dr/d\tau =0$, we arrived at
\begin{align}
E^{2}(r^{2}+a^{2}+l^{2})^{2}+a^{2}L^{2}-2aEL(r^{2}+a^{2}+l^{2})-\left(\lambda +m^{2}r^{2}\right)\Delta -\Delta \left(L^{2}+a^{2}E^{2}-2aEL\right)=0.
\end{align}
Choosing $\Delta=r^2-2 M r$, the above algebraic equation can be reduced to,
\begin{align}
E^{2}\left(r^{4}+a^4+l^4+r^2 a^2+2 r^2 l^2 +2 a^2 l^2 +2 M r a^2\right)-L^{2}\left(r^{2}-2Mr-a^{2}\right)-2aEL(a^{2}+l^2+2Mr) \nonumber \\
-\Delta \left(m^{2} r^{2}+\lambda\right)=0.
\label{eq:on_shell}
\end{align}
From which one can solve for the energy per unit mass $\tilde{E}$ in terms of the angular momentum per unit mass $\tilde{L}$, Carter constant per unit mass $\tilde{\lambda}$ and radius $r$ as, 
\begin{align}
\tilde{E}=\left\{(r^{2}+l^{2})^2+a^4+a^2 (r^2+2 M r+2 l^2)\right\}^{-1}&\Big[ a\left(a^{2}+l^{2}+2Mr\right)\tilde{L}
\pm\sqrt{(r^{2}-2Mr)}\Big\{\Big(r^{2}+\tilde{\lambda} \Big)\Bigl(a^{4}+
\nonumber
\\
&(r^{2}+l^{2})^2+a^{2} (r^{2}+2 M r+2 l^{2})\Bigr)+\tilde{L}^{2}(r^2+l^2)^2\Big\}^{1/2}
\end{align}
From the above equation, we only consider the expression with \enquote*{+} sign. This is because, in the $ r \rightarrow \infty$ limit, only the positive sign produces $\tilde{E}=1$, while the other gives $\tilde{E}=-1$. So, if we consider a timelike particle arriving from spatial infinity, only the positive sign suits the present analysis. Along identical lines, the expression of angular momentum can also be derived from \ref{eq:on_shell} in terms of energy and radial distance, 
\begin{align}
\tilde{L} &= &\left\{a^2-(r^2-2 M r)\right\}^{-1}\Big[a\tilde{E}\left(a^{2}+l^{2}+2Mr\right)
\mp\sqrt{(r^{2}-2Mr)}\Big\{\tilde{E}^2(r^2+l^2)^2-(r^2-2 M r-a^2)\nonumber 
\\
& & (r^2+\tilde{\lambda})\Big\}^{1/2}\Big]
\label{eq:angular_on_Shell}
\end{align}
In this case, both the \enquote*{$\mp$} sign are allowed and they corresponds to co-rotating and counter-rotating orbits respectively. Following an identical pathway as for the massive particle, the angular momentum for the massless particle can be also be written down, which takes the following form, 
\begin{align}
L=E\left\{a^2-(r^2-2 M r)\right\}^{-1}&\Big[a\left(a^{2}+l^{2}+2Mr\right)
\mp\sqrt{(r^{2}-2Mr)}\Bigl\{\Big(r^2+l^2 \Big)^2-\dfrac{\lambda}{E^2} \Big(r^2-2 M r -a^2\Big)\Bigr\}^{1/2}\Big]~.
\end{align}
As evident this is directly proportional to energy. Given the above ingredients one consider a variant of the Penrose process, where a massive particle arrives from infinity at the equatorial plane and then decays to two massless particles. One of them, with negative energy, falls into the black hole while the other one escapes to infinity. Since the decay process  is assumed to be taking place on the equatorial plane, the Carter constant for the massive particle takes the value $\tilde{\lambda}=l^2$, while the produced photons have the following values of Carter constants, $\lambda_1=0=\lambda_2$. Note that this does not require the timelike geodesic to be confined on the equatorial plane, rather it is expected that the massive particle has arrived from a different angular plane at infinity, as $\ddot{\theta} \neq 0$. To simplify the process further, we will assume that the produced photons along with the decaying massive particle have vanishing radial and angular momentum.

If we assume that the massive particle coming from infinity has vanishing initial velocity, it follows that its energy is given by $E=m$. As this particle subsequently breaks up into two parts, the amount of energy extracted in the process becomes, 
\begin{align}
E_{\rm ext}=\frac{1}{2}\left[\sqrt{\frac{(a^{2}+2Mr+l^{2})}{r^{2}+l^{2}}}-1\right]
\end{align}
The amount of extracted energy will be higher if the decay into two massless particle happens close to the event horizon. Thus in the $r\rightarrow 2M$ limit, we end up getting the following expression for extracted energy
\begin{align}
E_{\rm ext}=\frac{1}{2}\left[\sqrt{1+\dfrac{a^2}{l^2+4 M^2}}-1\right]
\end{align}
It is interesting to note that by fixing the \N\ charge $l$ to its minimum value, i.e., $l=a$, there is no upper bound on the rotation parameter $a$, while the electric charge parameter identically vanishes. This way, extracted energy become
\begin{align}
E_{\rm ext}=\frac{1}{2}\left[\sqrt{1+\dfrac{a^2}{a^2+4 M^2}}-1\right]
\end{align}
For a large momentum parameter, $E_{\rm ext}$ reaches a similar bound as the Kerr black hole, otherwise it is always less than that. This clearly suggests that the original Penrose process in a \KNN~spacetime is less efficient than its Kerr counterpart.  

Having discussed the maximum amount of energy extracted from a variant of the original Penrose process in a \KNN~spacetime, let us now provide an upper bound on the extracted energy in the form of Wald inequality \cite{Wald:1974kya,Bardeen:1972fi}. This inequality explicitly depends on the geometry of the spacetime as well as the velocity components of the fragments. If a particle with initial energy \enquote*{E} and four velocity $\mathcal{U}^a$ breaks into two parts and one of them having negative energy falls into the event horizon, while the other escapes to infinity extracts energy from the black hole. Considering the energy of the particle falling into the horizon as $\mathcal{E}$ with spatial velocity $\boldsymbol{\mathcal{V}}$, the Lorentz factor is given as, 
\begin{equation}
\gamma=\dfrac{1}{\sqrt{1-|\boldsymbol{\mathcal{V}}|^2}}~.
\end{equation}
Thus following \cite{chandrasekhar1998mathematical} one immediately arrives at the following constraint on the energy, 
\begin{equation}
\gamma \left(E-|\boldsymbol{\mathcal{V}}|(E^2+g_{\rm tt})^{1/2}\right) < \mathcal{E} < \gamma \left(E-|\boldsymbol{\mathcal{V}}|(E^2+g_{\rm tt})^{1/2}\right)~.
\end{equation} 
(Note the difference in sign of $g_{\rm tt}$ from Ref. \cite{chandrasekhar1998mathematical} due to opposite sign convention) For any process of energy extraction to take place, we must have $\mathcal{E}<0$, which suggests 
\begin{equation}
|\boldsymbol{\mathcal{V}}| >\dfrac{E}{(E^2+g_{\rm tt})^{1/2}}=\dfrac{1}{(1+E^{-2}g_{\rm tt})^{1/2}}.
\label{Min_Vel}
\end{equation}
Thus for a given initial particle of energy $E$, the velocity of the in-falling particle will be minimum (implying maximum energy extraction by the out-going particle) whenever $g_{\rm tt}$ is maximum, since it appears in the denominator. In the present context for \KNN\ black hole the \enquote*{$\text{tt}$} component of the metric on the equatorial plane (assuming that the decay is happening on the equatorial plane) is given as
\begin{equation}
g_{\rm tt}=\dfrac{a^2-\Delta}{r^2+l^2}.
\label{eq:g_tt}
\end{equation}
As evident from the above expression $g_{\rm tt}$ will attain its maximum possible value when $\Delta$ vanishes, i.e., on the event horizon, where its value is given by, 
\begin{equation}
g^{\rm max}_{\rm tt}=\dfrac{a^2}{4 M^2+l^2}.
\label{eq:g_tt}
\end{equation}
Thus given \ref{eq:g_tt} one can easily determine the minimum value of the velocity $\boldsymbol{\mathcal{V}}_{\rm min}$. This will certainly depend on the specific angular momentum $a/M$, following which we have presented the quantity $|\boldsymbol{\mathcal{V}}_{\rm min}|$ in \ref{KNN_Tab_01} for a several choices of the specific angular momentum. 
\begin{table}[htp]
\begin{center}
\caption{The numerical value of the minimum velocity $|\boldsymbol{\mathcal{V}}_{\rm min}|$ introduced above is of importance to the energy extraction process. In this table we have provided numerical estimates for the minimum velocity $|\boldsymbol{\mathcal{V}}_{\rm min}|$ with $l=a$ and various choices of the ratio $a/M$. As evident for larger values of $a/M$, the minimum velocity decreases from being unity. }
\label{KNN_Tab_01}
\vskip 2mm
\begin{tabular}{ c c } 
\hline
\multicolumn{1}{p{3cm}}{\centering angular \\ momentum \\ $(a/M)$}&
 $|\boldsymbol{\mathcal{V}}_{\rm min}|$  \\ [1ex] 
 \hline\hline
  1.0 & 0.912871 \\ 
 \hline
   3.0  & 0.768706  \\
 \hline
   5.0  & 0.732828 \\
  \hline
   7.0   & 0.720838 \\
  \hline
   9.0  & 0.715575 \\
 \hline
 \hline
\end{tabular}
\end{center}
\end{table}
As evident from \ref{KNN_Tab_01}, the minimum velocity decreases with increase in the value of the specific rotation parameter, but still this requires the velocity of the in-falling particle to be $\sim 0.72c$. Contrasting this with the case of Kerr black, where $\boldsymbol{\mathcal{V}}_{\rm min}\sim 0.5c$, we observe that energy extraction through Penrose process is more difficult in the context of \KNN\ spacetime.  
\subsection{Ba\~{n}ados-Silk-West Process}

It is recently proposed by Ba\~{n}ados et. al. that the collisional energy between two particles computed in the center of mass frame, $E_{\rm cm}$, can diverge in a rotating spacetime \cite{Banados:2009pr}. Since then, this proposal has been investigated in many aspects along with different models, which only strengthened its validity as a more general phenomenon 
\cite{Tanatarov:2014fra,Harada:2010yv}. In the present purpose, we investigate the same in the \KNN~spacetime with $\Delta=r^2-2Mr$. Without going into details of the analysis, the computed energy in the center of mass frame, namely $E_{\rm cm}$ turns out to be, 
\begin{align}\label{BSW}
E_{\rm cm}^{2}&=\frac{2m_{0}^{2}}{(r^{2}+l^{2})}\Bigg[\left\{(r^{2}-a^{2}+l^{2})-L_{1}L_{2}+a(L_{1}+L_{2})\right\}
\nonumber
\\
&+\frac{1}{2}\frac{\left\{(r^{2}+a^{2}+l^{2})-aL_{2}\right\}}{\left\{(r^{2}+a^{2}+l^{2})-aL_{1}\right\}} \left\{(r^{2}+a^{2}+l^{2})-2aL_{1}+L_{1}^{2}\right\}
\nonumber
\\
&+\frac{1}{2}\frac{\left\{(r^{2}+a^{2}+l^{2})-aL_{1}\right\}}{\left\{(r^{2}+a^{2}+l^{2})-aL_{2}\right\}} \left\{(r^{2}+a^{2}+l^{2})-2aL_{2}+L_{2}^{2}\right\}\Bigg]
\end{align}
It is straightforward to note that $E_{\rm cm}$ diverges, whenever one of the colliding particle has an angular momentum $L$, equal to $L_{1,2}\equiv(r_{\rm H}^2+a^2+l^2)/a$. We would like to emphasize that in our case, the black hole horizon is located at $r_{\rm H}=2M$, independent of the rotation parameter or the NUT charge. Thus, unlike the case for Kerr black hole, in this situation the spectrum of angular momentum related to divergent center-of-mass energy, $E_{\rm cm}$ is wider, which may lead to not-so-rare occurrence of this ultra-high energy particle accelerator in the context of \KNN\ black hole. 
\subsection{Superradiance in \KNN\ spacetime}

Superradiance is another way of extracting energy from a rotating object originally proposed by Zel'dovich in the early seventies. He suggested that in a particular limit, the amplitude of the reflected wave scattered by a rotating object can be larger than the amplitude of the incident wave. However, this rotating object has to have a well defined boundary and Zel'dovich had conducted his experiment with a rotating cylinder \cite{zel1971generation}. Afterwards, the idea to include the model of a black hole spacetime to explain the \SP~was investigated in Refs. \cite{Bekenstein:1973mi,starobinskil1973amplification,misner1972stability} and investigated by many others \cite{Richartz:2013hza,Eskin:2015ssa,Wagh:1986cz,Ganchev:2016zag}. In the present context, we use the \KNN~spacetime with $\Delta=r^2-2 M r$ and explore the possibilities of energy extraction via \SP. 

To complete the task, we shall assume a scalar field $\Phi$ defined as,
\begin{align}
\square \Phi = \frac{1}{\sqrt{-g}}\partial _{\mu}\left(\sqrt{-g}g^{\mu \nu}\partial _{\nu}\Phi \right)=0
\end{align}
We consider that the incident wave is scattered by the event horizon and a part of this gets transmitted across the horizon, while the other is reflected and travels to spatial infinity. 
Following the standard text books formalism with assuming an ansatz given as,
\begin{align}\label{super01}
\Phi=e^{-i\omega t}e^{im\phi}\Theta (\theta)R(r)
\end{align}
we arrive at the flux of energy through the horizon, 
\begin{align}\label{super02}
\frac{dE}{dt}&=\omega \left(\omega-m\Omega _{\rm H}\right)\left(r_{\rm H}^{2}+a^{2}+l^{2}\right)\int d\theta d\phi \rho_{\rm H}^{2}\sin ^{2}\theta \frac{\Theta ^{2}}{\rho_{\rm H}^{2}} 
\nonumber
\\
&= \omega \left(\omega-m\Omega _{\rm H}\right)(2 M r_{\rm H}+a^2+l^2)~\textrm{constant}
\end{align}
At this outset we would like to make a few remarks, which will help to understand the implications of the above expression for energy loss due to superradiance. Firstly, with $\omega < m \Omega_{\rm H}$, the energy flux has a negative sign which essentially indicates that there is a nonzero amount of energy carried out to infinity and \SP~is possible. On the other hand, for $\omega > m \Omega_{\rm H}$, the rate of energy loss is positive and hence \SP\ does not take place. Secondly, in a Kerr and Kerr-Newman spacetime, the pre-factor of the above equation becomes $2Mr_{\rm H}$ and $2Mr_{\rm H}-Q^2$ respectively, which is smaller compared to the \KNN\ black hole discussed in the present context. However the value of $\Omega _{\rm H}$ is less in the context of \KNN\ black hole and hence the frequency range upto which super-radiance can occur is smaller. Thus the total energy radiated by the superradiant modes in the present \KNN\ black hole is smaller compared to the Kerr or more general \KNN\ black holes. 
\section{Astrophysical Signaturesof \KNN\ black hole: Quasi-Periodic Oscillations}
\label{KNN_QPO}

Having addressed most of the energy extraction processes associated with the \KNN\ black hole, for completeness let us also mention one astrophysical implication of the same. There exist several possibilities to be explored, including luminosity from a black hole in the \KNN\ family and iron-line spectroscopy of the radiation originating from accretion disc around a \KNN\ black hole. However, in this work we will concentrate on the quasi-periodic oscillation from a accreting black hole which is described by the \KNN\ spacetime. Quasi-Periodic Oscillations (henceforth QPO) are related to very fast flux variability associated with matter accreting onto a black hole located very close to the innermost stable circular orbit. These QPOs are essentially believed to probe the geodesic motion of a particle in the strong field regime \cite{Stella:1998mq,Stella:1999sj,Casella:2005vy}. Till now, there are several black hole candidates, varying from stellar mass black hole to a supermassive one \cite{vanderKlis:2000ca,Motta:2013wga} for which such QPOs were observed. The frequencies of these QPOs are related to the fundamental frequencies associated with the motion of accreting matter in the strong gravity regime \cite{Stella:1998mq,Bambi:2013sha,Bambi:2012pa} and may probe the presence of a NUT charge if the background is given by \KNN\ spacetime. Further using the relativistic precession model, one can indeed predict correct values of black hole hairs (namely mass and angular momentum in the context of Kerr black hole) starting from the observations of QPOs \cite{Motta:2013wga,McClintock:2011zq}. Thus it is important to ask, whether it is possible to test the NUT hair as well using QPOs. In this section we will describe how the frequencies of QPOs depend on the NUT charge, while a numerical estimation by invoking real measurements will be done elsewhere \cite{Chakraborty:Prep}.  

We will work exclusively within the framework of relativistic precession model, which can explain the low frequency QPOs appearing in low mass X-ray binaries. In this model, the observed frequencies are very much related to the epicyclic frequencies associated with the radial, azimuthal and vertical motion. In the present context of \KNN\ black hole, due to its stationary and axisymmetric nature we can have two conserved quantities, namely the energy $E$ and angular momentum $L$. Using these two we can construct the following ratio $\ell=L/E$, in terms of which the effective potential takes the following form,
\begin{align}
V_{\rm eff}(r,\theta)&=g^{tt}-2\ell g^{t\phi}+\ell ^{2}g^{\phi \phi}
\nonumber
\\
&=\frac{1}{\Delta \rho ^{2}\sin ^{2}\theta}\Big[\Delta \left(P-\ell\right)^{2}
-\sin ^{2}\theta \left\{\left(r^{2}+a^{2}+l^{2}\right)-a\ell\right\}^{2}\Big]
\end{align}
Using the above expression for effective potential the radial equation of motion in the equatorial plane reads,
\begin{align}
\dot{r}^{2}&=-\frac{1}{g_{rr}}\left\{1+E^{2}V_{\rm eff}(r,\frac{\pi}{2})\right\}
\nonumber
\\
&=-\frac{\Delta}{r^{2}+l^{2}}\left[1+\frac{E^{2}}{\Delta\left(r^{2}+l^{2}\right)}
\left\{\Delta \left(a-\ell\right)^{2}-\left(r^{2}+a^{2}+l^{2}-a\ell\right)^{2} \right\} \right]
\end{align}
The azimuthal epicyclic frequency can be expressed as $\nu_{\phi}=\Omega/2\pi$, where $\Omega$ corresponds to the angular velocity of circular motion in the spacetime on equatorial plane. The angular velocity $\Omega$ can be solved starting from the following quadratic equation,
\begin{align}
\partial _{r}g_{tt}+2\left(\partial _{r}g_{t\phi}\right)\Omega +\left(\partial _{r}g_{\phi \phi}\right)\Omega ^{2}=0
\end{align}
which in the present context reads,
\begin{align}
-\Big\{2Mr^{2}-2Ml^{2}&+2r\left(a^{2}+l^{2}\right)\Big\}
-2\left\{2Mar^{2}+2Mal^{2}-2ar\left(a^{2}+l^{2}\right)\right\}\Omega 
\nonumber
\\
&+\left\{2r\left(r^{2}+l^{2}\right)^{2}+2a^{2}\left(M-r\right)\left(r^{2}+l^{2}\right)-2a^{2}r\left(a^{2}-r^{2}+2Mr\right) \right\}\Omega ^{2}=0
\end{align}
which can be solved to yield the azimuthal frequency $\nu_{\phi}$ through $\Omega$.

The radial and vertical epicyclic frequencies are obtained by considering small perturbations from the circular motion, such that $r=r_{c}+\delta r$ and $\theta=(\pi/2)+\delta \theta$, where $r_{c}$ is the radius of circular orbit. The perturbations are assumed to be oscillating such that, $\delta r\sim \exp(2\pi i \nu_{r}t)$ and $\delta \theta \sim \exp(2\pi i \nu_{\theta}t)$, where $\nu_{r}$ and $\nu_{\theta}$ are the desired frequencies. It is indeed possible to express these frequencies in terms of the metric elements of \KNN\ spacetime, such that,  
\begin{align}
\nu_{r}^{2}&=\frac{\left(g_{tt}+\Omega g_{t\phi}\right)^{2}}{2\left(2\pi\right)^{2}g_{rr}}
\frac{\partial ^{2}V_{\rm eff}}{\partial r^{2}}\left(r_{c},\frac{\pi}{2}\right)
\nonumber
\\
&=\frac{1}{2\left(2\pi\right)^{2}}\frac{\Delta}{r^{2}+l^{2}}
\left(-\frac{\Delta-a^{2}}{r^{2}+l^{2}}-\Omega\frac{a\left(a^{2}+l^{2}+2Mr\right)}{r^{2}+l^{2}}\right)^{2}
\frac{1}{\Delta ^{3}\left(r^{2}+l^{2}\right)^{3}}
\nonumber
\\
&\times \Bigg[-8r^{2}\Delta ^{2}\left(r^{2}+l^{2}\right)^{2}+\Delta ^{3}\left(a-\ell\right)^{2}\left(-10r^{2}-2l^{2}\right)
+\left(r^{2}+a^{2}+l^{2}-a\ell \right)\Big\{-4\Delta ^{2}\left(r^{2}+l^{2}\right)^{2} 
\nonumber
\\
&+8r\Delta \partial _{r}\Delta \left(r^{2}+l^{2}\right)^{2}+16\Delta ^{2}r^{2}\left(r^{2}+l^{2}\right) \Big\}
+\left(r^{2}+a^{2}+l^{2}-a\ell \right)^{2}\Big\{-8r^{2}\Delta ^{2}+2\Delta ^{2}\left(r^{2}+l^{2}\right)
\nonumber
\\
&-4r\Delta \partial _{r}\Delta \left(r^{2}+l^{2}\right)+2\left(r^{2}+l^{2}\right)^{2}\left(\Delta -\partial _{r}\Delta ^{2} \right)\Big\}\Bigg]
\\
\nu_{\theta}^{2}&=\frac{\left(g_{tt}+\Omega g_{t\phi}\right)^{2}}{2\left(2\pi\right)^{2}g_{\theta \theta}}
\frac{\partial ^{2}V_{\rm eff}}{\partial \theta^{2}}\left(r_{c},\frac{\pi}{2}\right)
\nonumber
\\
&=\frac{1}{2\left(2\pi\right)^{2}}\frac{1}{\left(r^{2}+l^{2}\right)}\left(-\frac{\Delta-a^{2}}{r^{2}+l^{2}}-\Omega\frac{a\left(a^{2}+l^{2}+2Mr\right)}{r^{2}+l^{2}}\right)^{2}
\frac{1}{\Delta}\frac{1}{\left(r^{2}+l^{2}\right)^{3}}\Bigg[\left(r^{2}+l^{2}\right)^{2}\Big\{8\Delta l^{2}-4a\Delta \left(a-\ell \right)
\nonumber
\\
&+2\left(r^{2}+a^{2}+l^{2}-a\ell \right)^{2}\Big\}+\left(r^{2}+l^{2}\right)\Big\{16al^{2}\Delta \left(a-\ell\right)
+2\left(r^{2}+l^{2}+a^{2}\right)
\nonumber
\\
&\times \left[\Delta \left(a-\ell \right)^{2}- \left(r^{2}+a^{2}+l^{2}-a\ell \right)^{2} \right]\Big\}
+8a^{2}l^{2}\Big\{\Delta  \left(a-\ell\right)^{2}- \left(r^{2}+a^{2}+l^{2}-a\ell \right)^{2}\Big\}\Bigg]
\end{align}
As evident both $\nu_{r}$ and $\nu_{\theta}$ depends explicitly on the NUT parameter $l$ and hence these epicyclic frequencies indeed inherits the NUT charge as an additional hair. Thus the QPO frequencies will be affected (or, modified) by the presence of NUT charge and may lead to some interesting bound on the same when compared to observational measurements. This provides one of the astrophysical grounds to test the existence of the NUT parameter on black holes within the the \KNN\ family (for an attempt in a similar direction, see \cite{Chakraborty:2017nfu}). Besides, the special structure of $\Delta$ appearing in the situation under consideration modifies the QPO frequencies in a significant manner, compared to the other members of the \KNN\ family. Furthermore, the fact that neither rotation nor NUT parameter is restricted within any upper boundary (i.e., they can both be greater than unity) suggests that a wider range of  parameter space can be accessed. This in principle can be useful to shed light on the possible existence of the NUT charge in nature in a more efficient way. 
\section{Thermodynamics of \KNN\ black hole}
\label{KNN_Thermo}

In this section, we will try to understand some thermodynamical aspects of the \KNN~black hole. The first in the list corresponds to the computation of area associated with the black hole horizon, which can be obtained by setting $\Delta \rightarrow 0$. This leads to,
\begin{align}
\textrm{Area}_{\rm H}&=\int _{0}^{\pi}d\theta \int _{0}^{2\pi}d\phi \sqrt{h_{\theta \theta}h_{\phi \phi}}
=\int _{0}^{\pi}d\theta \int _{0}^{2\pi}d\phi \sin \theta \left(r_{H}^{2}+a^{2}+l^{2}\right),
\nonumber
\\
&=4\pi \left(r_{H}^{2}+a^{2}+l^{2}\right)~.
\end{align}
From the above expression, the entropy of the horizon can be written as, 
\begin{align}
S_{\rm H}=\dfrac{\textrm{Area}_{\rm H}}{4}=\pi \left(r_{H}^{2}+a^{2}+l^{2}\right)~.
\end{align}
The other most important thing corresponds to the surface gravity, or the temperature associated with the black hole \cite{poisson2004relativist}. This originates from the Killing vector field,
\begin{align}
\xi ^{\mu}=t^{\mu}+\Omega _{\rm H}\phi ^{\mu}
\end{align}
where, $t^{\mu}=(\partial /\partial t)^{\mu}=(1,0,0,0)$ is the Killing vector ensuring stationarity and $\phi ^{\mu}=(\partial /\partial \phi)^{\mu}=(0,0,0,1)$ is the Killing vector from axi-symmetry. Thus norm of the vector $\xi ^{\mu}$ yields,
\begin{align}
\xi ^{\mu}\xi _{\mu}&=g_{\mu \nu}\left(t^{\mu}+\Omega _{\rm H}\phi ^{\mu}\right)
\left(t^{\nu}+\Omega _{\rm H}\phi ^{\nu}\right)
=g_{tt}+2\Omega _{\rm H}g_{t\phi}+\Omega _{\rm H}^{2}g_{\phi \phi}
\end{align}
Substituting for the metric elements as one approaches the horizon along with the angular velocity $\Omega _{\rm H}$ at the horizon, the norm turns out to be,
\begin{align}\label{eq:ADM}
\xi ^{\mu}\xi _{\mu}=-\frac{\Delta \sin ^{2}\theta \left\{aP-\left(r^{2}+a^{2}+l^{2}\right)\right\}^{2}}{\rho ^{2}\left\{\sin ^{2}\theta \left(r^{2}+a^{2}+l^{2}\right)^{2}-\Delta P^{2} \right\}}
\end{align}
where the symbols have been introduced earlier, see e.g., \ref{def_eq}. Since we are interested in the $\Delta \rightarrow 0$ limit, the derivative of \ref{eq:ADM} will be proportional to the radial derivative of $\Delta$. Thus the surface gravity can be read off from the result $\nabla _{\alpha}(\xi ^{\mu}\xi _{\mu})=-2\kappa \xi _{\alpha}$, as,
\begin{align}
\kappa = \frac{r_{\rm H}-M}{\left(r_{\rm H}^{2}+a^{2}+l^{2}\right)}
\end{align}
For $\Delta=r_{\rm H}^{2}-2Mr_{\rm H}=0$, we obtain the associated temperature to be,
\begin{align}
T=\frac{\kappa}{2\pi} = \frac{M}{2\pi\left(2Mr_{\rm H}+a^{2}+l^{2}\right)}
\end{align}
It is interesting to note one particular difference from the usual Kerr black hole regarding the extremal condition, i.e, at $a=M$ limit. Even though surface gravity or temperature identically vanishes in this limit for a Kerr spacetime, in the present context with $\Delta=r^2-2Mr$, both of them have nonzero contribution. 

\section{Concluding remarks}\label{KNN_Conc}

The duality between gravitational mass and NUT charge makes the \KNN\ spacetime an interesting testbed to understand gravitational physics. It has been shown in \cite{Dadhich:2001sz,Turakulov:2001jc} that Kerr-NUT metric is invariant under the duality transformation: $M \leftrightarrow il$, $r \leftrightarrow i(l + a \cos\theta)$, exhibiting duality between gravoelectric ($M$) and gravomagnetic ($l$) charge \cite{LyndenBell:1996xj},  and correspondingly between radial and angular coordinates. In the \KNN~metric, ($M, Q$) are gravoelectric charges (which are purely coloumbic in nature in the sense that they appear only in $\Delta$) while ($l, a$) are gravomagnetic, which in addition to $\Delta$ (since energy in any form must gravitate) also appear in the metric defining the geometric symmetry of the spacetime.

In generic situations the \KNN\ spacetime inherits two horizons and a \emph{timelike} singularity. However, for a particular choice of the charge parameter, namely, $Q^{2}+a^{2}=l^{2}$, where $l$ and $a$ are the NUT charge and the black hole rotation parameter respectively, the horizon sits at $r=2M$. This particular relation between the black hole hairs, namely the electric charge $Q$, NUT charge $l$ and rotation parameter $a$ ensures that the amount of repulsion offered by $a$ and $Q$ is being exactly balanced by the attraction due to the NUT charge and hence as a consequence the horizon is located at a position as if none of these hairs are present. Note that the horizon can appear at $r=2M$ even for  $l=a, Q=0$, a particular case of the Kerr-NUT spacetime, where the two magnetic hairs $l$ and $a$ are equal. Interestingly, for $l=a$ the curvature singularity is spacelike in stark contrast to the generic \KNN\ spacetimes. The radical alteration in casual structure due to the above choice is intriguing and has remained unnoticed in the literature. On the other hand, for $l^2 > a^2$ it follows that $ r^2 + (l + a \cos\theta)^2 \neq 0$ for any real value of $r$ and hence remarkably the solution presents a black hole solution free of any curvature singularity. The above prescription squarely balances coloumbic gravitational effects of charge, rotation and NUT parameter leaving mass alone to determine the horizon and the nature of curvature singularity. However $l$ and $a$ have not been fully eliminated as they also occur in the metric in their magnetic role as in $\rho^2$. It is their magnetic role which is very interesting and would require a separate detailed and deeper investigation. However the key fact derived here corresponds to the result that when NUT parameter is dominant over rotation, singularity is avoided. That means magnetic contribution of $l$ and $a$ dominates over coloumbic contribution due to mass. On the other hand, when $l \leq a$, mass dominates and a spacelike singularity arises. Since location of horizon is free of NUT, rotation and charge parameters, they could have any value, even greater than unity. For the generic \KNN\ family, one of the interesting extremal configuration could be equality of gravoelectric and gravomagnetic charges, i.e., $M=Q, l=a$, separately. It should be interesting to study this kind of extremal \KNN\ black hole in an extensive scale.  

Given all these distinguishing features associated with the above black hole spacetime, we have studied the trajectories of massive as well as massless particles in this spacetime. As we have explicitly demonstrated, in general circular geodesics cannot be confined to the equatorial plane. Following which we have determined the conditions on the angle $\theta$ for which circular geodesics are possible and it turns out that non-zero Carter constant plays a very important role in this respect. Besides we have also discussed geodesic motion on non-equatorial planes to understand its physical properties in a more comprehensive manner. In particular, we have studied the photon circular orbits as well as circular orbits for massive particles on a fixed $\theta=\textrm{constant}$ plane and have determined the possible conditions on the energy for their existence. The results obtained thereafter explicitly demonstrate the departure of the present context from the usual \KNN\ scenario, e.g., the innermost stable circular orbits are located at a completely different angular plane and also at different location compared to the corresponding radius in Schwarzschild spacetime. This may have interesting astrophysical implications, e.g., this will affect the structure of accretion disk around the black hole, which in turn will affect the observed luminosity from the accretion disk. Besides the above, we have also studied various energy extraction processes in this spacetime. It turns out that in both the Penrose process and super-radiance the amount of energy extracted is less in comparison to the corresponding situation with Kerr black hole. On the other hand, in the black hole spacetime under consideration, the center of mass energy of a pair of colliding particles can be very large (the Banados-Silk-West effect) for a much wider class of angular momentum of the incoming particles. This is also in sharp contrast with the corresponding scenario for Kerr spacetime. In addition, the fundamental QPO frequencies for a geodesic trajectory orbiting in nearly circular orbits on the equatorial plane are also carried out in detail pointing out its astrophysical significance. It is also stressed that the domain of the black hole parameters $l$ and $a$ leads to a much large parameter space, which could be spanned to look for any signature of the NUT charge in astrophysical scenarios. This may provide a better scope for estimating the parameters in order to match with the observational data. Further we have also commented on the thermodynamical aspects, in which case unlike the general \KNN\ spacetime, the black hole temperature does not vanish for any parameter space of the NUT charge and the rotation parameter. We would like to emphasize that the results derived in this work are qualitatively different from the results presented in the earlier literature, see \cite{Griffiths:2009dfa,Grenzebach:2014fha,Johnson:2014pwa}. Since the case $\ell=a$ with $Q=0$ has not been studied extensively earlier in the literature, the results presented in this work has possibly shed some light into this parameter space of the \KNN\ solution and has filled a gap in the literature. However in certain arenas, e.g., study of photon circular orbits, one can use the results derived in \cite{Grenzebach:2014fha} to immediately observe that our claim regarding non-equatorial motion are in direct consonance with earlier literatures. This provides yet another verification of our result presented in this work.

Finally, we would like to point out that for this particular relation among black hole hairs, namely, $Q^{2}+a^{2}=l^{2}$, the ratio $(a/M)$ is completely free and can even take values larger than unity, while at the same time if $l>a$, the black hole will be free of any curvature singularity. Thus any observational evidence that indicates a possibility of having a super-rotating black hole with $(a/M)>1$, need not necessarily be a signature of naked singularity but instead it could as well be a case of black hole having a NUT charge. This, as well as all other features mentioned above would indeed make a good case for studying the role of NUT parameter in high energy astrophysical setting and phenomena.
\section*{Acknowledgement} 

Research of S.C. is supported by the INSPIRE Faculty Fellowship (Reg. No. DST/INSPIRE/04/2018/000893) from Department of Science and Technology, Government of India. S.C. and N.D. acknowledge the warm hospitality provided by the Albert Einstein Institute, Golm, Germany. Finally, S.C. and S.M. would like to thank the Inter-University Centre for Astronomy and Astrophysics (IUCAA), Pune, India where parts of this work were carried out during short visits. 

\bibliography{References}
\bibliographystyle{./utphys1}

\end{document}